\documentclass[acmsmall]{acmart}
\usepackage{blindtext}
\usepackage{cleveref}
\usepackage{graphicx}
\usepackage{hyperref}
\usepackage{listings}
\usepackage{subfiles}
\usepackage{todonotes}
\usepackage{subcaption}
\usepackage{multirow}
\usepackage{mathpartir}
\usepackage{amsmath}

\newcommand{\F}[1]{\text{\textsf{#1}}\xspace}
\newcommand{\K}[1]{\text{\textbf{#1}}\xspace}
\newcommand{\name}{OpTuner\xspace}

\newcommand{\float}[2]{\ensuremath{#1 \cdot 10^{#2}}}

\newcommand{\nSpecThreshold} {$0.96$\xspace}
\newcommand{\nBenchmarks} {$37$\xspace}
\newcommand{\nBenchmarkUseSites} {$89$\xspace}
\newcommand{\nBenchmarkSafeSpeedup} {$107\%$\xspace}
\newcommand{\nBenchmarkLargeSpeedup} {$438\%$\xspace}
\newcommand{\nBenchmarkAvgMaxSpeedup} {$244\%$\xspace}
\newcommand{\nDetailedOneSpeedup} {$17\%$\xspace}
\newcommand{\nDetailedOneErrorGlibc} {$\float{6.47}{-16}$\xspace}
\newcommand{\nDetailedOneErrorChosen} {$\float{8.99}{-15}$\xspace}
\newcommand{\nDetailedTwoSpeedup} {$79\%$\xspace}
\newcommand{\nDetailedTwoErrorGlibc} {$\float{2.25}{-11}$\xspace}
\newcommand{\nDetailedTwoErrorChosen} {$\float{8.65}{-11}$\xspace}
\newcommand{\nDetailedThreeSpeedup} {$205\%$\xspace}
\newcommand{\nDetailedThreeErrorGlibc} {$\float{4.61}{-15}$\xspace}
\newcommand{\nDetailedThreeErrorChosen} {$\float{4.99}{-6}$\xspace}
\newcommand{\nOptunerImpls} {$216$\xspace}
\newcommand{\nOptunerImplErrorOrders} {$15$\xspace}
\newcommand{\nOptunerImplSpeed} {$58\times$\xspace}
\newcommand{\nCosImpls} {$45$\xspace}
\newcommand{\nSinImpls} {$49$\xspace}
\newcommand{\nScarySinError} {$\float{3.87}{-6}$\xspace}
\newcommand{\nScarySinTime} {$3.85$ nanoseconds\xspace}
\newcommand{\nPovRayTableImplTime} {$3.19$ nanoseconds\xspace}
\newcommand{\nPovProgConfigs} {$219$\xspace}
\newcommand{\nPovProgCrlibmSlowdown} {$5.07\times$\xspace}
\newcommand{\nPovProgCrlibmAccuracyIncrease} {$1.53\times$\xspace}
\newcommand{\nPovProgTableSpeedup} {$2.15\times$\xspace}

\newcommand{\nPovProgLosslessAccuracyDecrease} {$\float{3.5}{6}\times$\xspace}
\newcommand{\nPovProgLosslessSpeedup} {$2.48\times$\xspace}
\newcommand{\nPovProgFastSpeedup} {$4.46\times$\xspace}
\newcommand{\nPovProgFastAccuracyDecrease} {$\float{3.72}{14}\times$\xspace}
\newcommand{\nPovProgFastRelativeError} {$0.259$\xspace}
\newcommand{\nPovProgOptunerTime} {$82$ minutes\xspace}

\newcommand{\nPovRaySatConfigs} {$78$\xspace}
\newcommand{\nPovRayLosslessSpeedup} {$9\%$\xspace}
\newcommand{\nPovRayPhotonsPercent} {$60\%$\xspace}

\newcommand{\nPovRayConstSpeedup} {$1.83\%$\xspace}

\newcommand{\nPovRayCharSpeedup} {$10.6\%$\xspace}
\newcommand{\nPovRayPossible} {$4.9$ million\xspace}

\newcommand{\nBenchImpls} {$1175$\xspace}
\newcommand{\nBenchDomainViolationRate} {$11.0\%$\xspace}

\newcommand{\nMaxTerms} {$16$\xspace}

\newcommand{\nTotalGenerationTime} {$2$ hours\xspace}
\newcommand{\nSollyaAvg} {$5$ minutes\xspace}
\newcommand{\nErrorVerrifyAvg} {$3$ minutes\xspace}
\newcommand{\nReductionErrorAvg} {$3$ minutes\xspace}
\newcommand{\nTimingInputs} {$163840$\xspace}
\newcommand{\nTimingElapsed} {$10$ seconds\xspace}
\newcommand{\nBenchmarksUnderThreeMinutes} {$28$\xspace}

\newcommand{\nBenchmarksUnderTenMinutes} {$33$\xspace}

\setcopyright{acmcopyright}
\copyrightyear{2018}
\acmYear{2018}
\acmDOI{10.1145/1122445.1122456}
\acmConference[PL'18]{ACM SIGPLAN Conference on Programming Languages}{January 01--03, 2018}{New York, NY, USA}
\acmYear{2018}
\acmISBN{}
\acmDOI{}
\startPage{1}
\setcopyright{none}


\begin{document}

\title{Faster Math Functions, Soundly}

\author{Ian Briggs}
\email{ibriggs@cs.utah.edu}
\affiliation{%
  \institution{University of Utah}
  \city{Salt Lake City}
  \state{UT}
  \country{USA}
}

\author{Pavel Panchekha}
\email{pavpan@cs.utah.edu}
\affiliation{%
  \institution{University of Utah}
  \city{Salt Lake City}
  \state{UT}
  \country{USA}
}

\begin{abstract}

Standard library implementations
  of functions like \F{sin} and \F{exp}
  optimize for accuracy, not speed,
  because they are intended for general-purpose use.
But applications tolerate inaccuracy
  from cancellation, rounding error, and singularities---%
  sometimes even very high error---%
  and many application
  could tolerate error in function implementations as well.
This raises an intriguing possibility:
  speeding up numerical code
  by tuning standard function implementations.

This paper thus introduces \name,
  an automatic method for selecting
  the best implementation of mathematical functions
  at each use site.
\name assembles dozens of implementations
  for the standard mathematical functions
  from across the speed-accuracy spectrum.
\name then uses error Taylor series and integer linear programming
  to compute optimal assignments
  of function implementation to use site
  and presents the user with a speed-accuracy Pareto curve
  they can use to speed up their code.
In a case study on the POV-Ray ray tracer,
  \name speeds up a critical computation,
  leading to a whole program speedup of \nPovRayLosslessSpeedup
  with no change in the program output
  (whereas human efforts result in slower code and lower-quality output).
On a broader study of \nBenchmarks standard benchmarks,
  \name matches \nOptunerImpls implementations
  to \nBenchmarkUseSites use sites
  and demonstrates speed-ups
  of \nBenchmarkSafeSpeedup for negligible decreases in accuracy
  and of up to \nBenchmarkLargeSpeedup for error-tolerant applications.

\end{abstract}

\maketitle

\section{Introduction}
\label{sec:introduction}

Floating-point arithmetic is foundational for scientific, engineering,
  and mathematical software.
This is because, while floating-point arithmetic is approximate,
  most applications tolerate minute errors~\cite{cern}.
In fact, a speed-accuracy trade-off
  is ever-present in numerical software engineering:
  mixed-and lower-precision floating-point~\cite{fptuner,precimonious,salsa,hifptuner},
  alternative numerical representations~\cite{bfloat16,posit,block-numbers},
  quantized or fixed-point arithmetic~\cite{fixed},
  rewriting~\cite{pherbie},
  and various forms of lossy compression~\cite{fpcompression1,fpcompression2}
  all promise faster but less accurate programs.
In each case,
  the challenge is helping the numerical software engineer
  apply the technique
  and explore the trade-offs available.

The implementation of library functions like \F{sin}, \F{exp}, or \F{log}
  is one instance of this ever-present speed-accuracy trade-off.
Traditional library implementations, such as those in GLibC,
  guarantee that all but maybe the last bit are correct.
This high accuracy comes at a cost:
  these traditional implementations tend to be slow.
Recent work shows that substantial speed-ups are possible%
  ~\cite{cern,metalibm,daisy-libm,vanover}
  if the accuracy requirement is relaxed.
But achieving that speed-up in practice is challenging,
  because it requires carefully selecting
  among alternate function implementations~\cite{intel,cern,amd,openlibm},
  and proving accuracy bounds for the tuned program.
All this requires manual effort and deep expertise,
  along with a substantial investment of time and effort.

We propose \name, a sound, automatic tool
  for exploring the speed-accuracy trade-offs of library function implementations.
For any floating-point expression,
  \name selects the best \F{exp}, \F{log}, \F{sin}, \F{cos}, or \F{tan}
  implementation to use for each call site in the program
  and presents the user with a speed-accuracy Pareto curve.
This Pareto curve
  shows only the best-in-class tuned implementations,
  condensing millions of possible configurations
  into a few dozen that the user can rapidly explore
  to speed up their computation.
Each point on the curve is also annotated
  with a rigorously-derived sound error bound
  allowing non-experts to understand
  its impact on the accuracy of their code.

\name's key insight
  is that error Taylor series, a state of the art technique
  for bounding floating-point error,
  can be extended to derive
  a \textit{linear error model}
  that predicts the error of the expression
  based on the error of the individual function implementations used.
This linear error model is combined
  with a simple, linear, cost model for expression run time
  to create an integer linear program
  whose discrete variables encode the choice
  of function implementation for each use site.
An off-the-shelf integer linear programming solver
  is then be used to rapidly search
  through millions of possible implementation choices
  to find the points along the speed-accuracy Pareto curve
  for the input expression.
The error and speed is then verified by timed execution and computation of a
  sound error bound before being
  presented to the user.

One of the benefits of this approach
  is that \name can perform optimizations
  too difficult or nit-picky for humans to perform.
We illustrate this
  by introducing custom implementations
  for \F{exp}, \F{log}, \F{sin}, \F{cos}, and \F{tan}
  that are only valid on a restricted range of inputs.
The restricted range allows
  the use of simplified range reduction methods
  that are much faster than a traditional implementation.
Using these new implementations requires proving
  that their input is within a certain range of values,
  even when taking into account
  the rounding error incurred computing that input.
We extend \name to automatically perform such proofs,
  and therefore transparently mix
  these range-restricted implementations with existing libraries
  to achieve even better speed-accuracy trade-offs.

We evaluate \name on \nBenchmarks standard benchmarks
  from the FPBench 2.0~\cite{fpbench} and Herbie 1.5 suites~\cite{herbie}.
\name tunes these benchmarks using
  \nOptunerImpls implementations
  of \F{sin}, \F{cos}, \F{tan}, \F{exp}, and \F{log},
  ranging in accuracy across \nOptunerImplErrorOrders different orders of magnitude
  and with speeds that vary by a factor of \nOptunerImplSpeed.
\name can provide a speedup of \nBenchmarkSafeSpeedup
  while maintaining high accuracy,
  and for error-tolerant applications
  the \name-optimized benchmarks
  are up to \nBenchmarkLargeSpeedup faster
  than ones that use the GLibC implementations.

To highlight the benefits of \name,
  we perform a case study with the POV-Ray ray tracer.
POV-Ray is a state-of-the-art CPU ray tracer
  and part of the SPEC 2017 benchmark collection.
We find that POV-Ray
  spends a substantial portion of its runtime
  inside calls to \F{sin} and \F{cos},
  and the POV-Ray developers
  maintain custom \F{sin} and \F{cos} implementations
  that are faster but less accurate
  in order to achieve acceptable speed.
We show that \name can automate this kind of optimization,
  achieving an end-to-end \nPovRayLosslessSpeedup speed-up
  with no loss in output quality.
This is both faster and higher quality
  than the POV-Ray developers' own efforts.
Moreover, other points on \name's speed-accuracy Pareto curve
  could be useful to the POV-Ray developers
  or even to users with complex geometries.

\medskip
\noindent
In summary, our main insight
  is that error Taylor series can be used
  to derive a \textit{linear error model}
  for the accuracy of a floating-point expression
  in terms of the function implementations it uses.
That allows us to construct \name, a tool with:
\begin{itemize}
\item Accuracy specifications for widely used mathematical libraries (\Cref{sec:implementations});
\item Automatically-derived linear cost models for error and runtime (\Cref{sec:error-and-cost-models});
\item An integer linear formulation of the implementation selection problem (\Cref{sec:optimization});
\item Extensions to handle function implementations with restricted input ranges (\Cref{sec:ranges});
\item Fast, range-restricted implementations of \F{exp}, \F{log}, \F{sin}, \F{cos}, and \F{tan}.
\end{itemize}
\Cref{sec:evaluation} demonstrates
  that by leveraging these components
  \name can dramatically speed up standard floating-point benchmarks
  with minimal loss of accuracy.

\section{The Big Idea in One Formula}
\label{sec:big-idea}

Floating-point arithmetic deterministically
  approximates real-number arithmetic.
The error of this approximation
  is given the rounding model
  $y(1 + \varepsilon) + \delta$,
  which formally states
\begin{equation}\label{rdmodel}
\forall\:x \in D,
\exists\:|\varepsilon_x| \le \varepsilon_f,
|\delta_x| \le \delta_f,
\quad
\tilde{f}(x) = f(x) (1 + \varepsilon_x) + \delta_x,
\end{equation}
In other words,
  a floating-point computation $\tilde{f}(x)$
  is equal to its true mathematical value $f(x)$,
  but with a relative error of $\varepsilon_x$
  (bounded by $\varepsilon_f$)
  and an absolute error of $\delta_x$
  (bounded by $\delta_f$).
Both $\varepsilon_x$ and $\delta_x$ are necessary
  to bound the error for both normal and subnormal numbers.%
\footnote{This is more important in our context
  than in traditional uses of error Taylor series,
  because $\delta_f$ can be quite large
  for some function implementations.}
The constants $\varepsilon_f$ and $\delta_f$,
  depend on the particular function implementation $\tilde{f}$
  and characterize its accuracy.

\subsection{Error Taylor Series}
\label{ssec:aspec}

\Cref{rdmodel} bounds the error of a single call to $f$,
  but a floating-point expressions calls multiple functions,
  and their errors interact to affect
  the \emph{overall} error of that expression.
Consider the composition $\tilde{f}(\tilde{g}(x))$.
Expanding according to \cref{rdmodel} yields
\begin{equation}\label{exfn}
  \tilde{f}(\tilde{g}(x)) = f(g(x)(1 + \underbrace{\varepsilon_1) + \delta_1}_{\text{from }\tilde{g}})(1 + \underbrace{\varepsilon_2) + \delta_2}_{\text{from }\tilde{f}}
\end{equation}
Here, the $\varepsilon_1$, $\delta_1$, $\varepsilon_2$, and $\delta_2$ terms
  are variables bounded by constants $\varepsilon_f$, $\delta_f$, $\varepsilon_g$, and $\delta_g$
  that characterize the error of $\tilde{f}$ and $\tilde{g}$.%
\footnote{The difference between, say, $\varepsilon_1$ and $\varepsilon_f$
  is subtle---the first represents the actual rounding error,
  while the second represents a worst-case error bound.
  The reader can ignore likely the difference without much harm.}
Error Taylor series are a state-of-the-art technique
  to bound the maximum error of this expression
  based on the insight
  that the $\varepsilon$ and $\delta$ parameters are always small.

The core idea is to expand \Cref{exfn}
  as a Taylor series in the $\varepsilon$s and $\delta$s:
\[
\tilde{f}(\tilde{g}(x)) =
f(g(x)) + f'(g(x))g(x)\varepsilon_1 + f'(g(x))\delta_1
+ f(g(x))\varepsilon_2 + \delta_2 + o(\varepsilon^2)
\]
The first term in this Taylor series represents the exact output,
  so the other terms represent the error.
Ignore for a moment the higher-order terms represented by $o(\varepsilon^2)$ as
  they are usually insignificant;
  the other four linear terms are called
  the \textit{first-order error} of the computation.

Since these four terms are linear
  in the $\varepsilon$ and $\delta$ terms,
  the maximum error occurs
  when each $\varepsilon$ and $\delta$
  has the largest possible magnitude
  and the same sign as the term it is multiplied by:
\begin{equation}\label{taylorform}
|\tilde{f}(\tilde{g}(x)) - f(g(x))| \le
\max_x \left( |f'(g(x))g(x)| \varepsilon_g \\ +
|f'(g(x))| \delta_g +
|f(g(x))| \varepsilon_f +
| 1 | \delta_f + o(\varepsilon^2) \right)
\end{equation}
Note that $\varepsilon_f$ and similar are constants, not variables,
  so bounding the maximum error of the linear terms
  just requires optimizing a complicated function of $x$,
  which can be done using a global nonlinear optimization package.
The higher-order terms can also be bounded,
  via Lagrange's theorem for the remainder of a Taylor series,
  again using global non-linear optimization.%
\footnote{
  Typically, the higher-order terms are too small to make a difference,
    but they are necessary to establish a sound error bound.}
Error Taylor series thus provide
  an elegant, rigorous, and relatively accurate way
  to bound the error of an arbitrary floating-point expression.
Recent work
  focuses on automating this idea~\cite{fptaylor}
  and scaling it larger programs~\cite{satire}.

\subsection{The Idea}

\name's key insight is to treat $\varepsilon_f$ and $\delta_f$
  not as constants but as variables.
By triangle inequality,
\begin{multline}\label{linearform}
  | \tilde{f}(\tilde{g}(x)) - f(g(x)) | \le
\max_x \left( |f'(g(x))g(x)| \varepsilon_g + |f'(g(x))| \delta_g + |f(g(x))|
\varepsilon_f + | 1 | \delta_f \right) \le \\
\underbrace{\left( \max_x |f'(g(x))g(x)| \right)}_{\text{constant }A_1} \varepsilon_g
+ \underbrace{\left( \max_x |f'(g(x))| \right)}_{\text{constant }B_1} \delta_g
+ \underbrace{\left( \max_x |f(g(x))| \right)}_{\text{constant }A_2} \varepsilon_f
+ \underbrace{\left( \max_x | 1 | \right)}_{\text{constant }B_2} \delta_f
\end{multline}
The coefficients $A_i$ and $B_i$
  in front of each $\varepsilon_f$ and $\delta_g$
  are numbers that can be directly computed
  using a global nonlinear optimizer.
This simple rearrangement thus
  converts the error Taylor series from a simple numeric bound
  to a linear \emph{error model}
  that gives the accuracy of the overall expression
  in terms of the errors of each individual function implementation.%
\footnote{
This derivation focuses on bounding absolute error,
  but an analogous error form can be derived for the relative error
  by divided each coefficient by $|f(g(x))|$.}
\name uses this linear error model in an integer linear program
  to search for optimal speed-accuracy trade-offs
  and then presents those trade-offs to the user.

\section{Case Study}
\label{sec:case_study}

Ray tracers are complex, numerically intense software
  that can run for hours or even days:
  the perfect target for \name.
Moreover, ray tracers are naturally tolerant of inaccuracy
  since they produce an image with only eight bits per color channel
  and the resulting images are often further compressed by image and video codecs.
We therefore conducted a case study applying \name to an expression extracted
  from POV-Ray~\cite{povray},
  a full-featured, widely-used, and mature ray tracer
  in continuous development since 1992.

Searching POV-Ray's source code for calls to \F{sin} and \F{cos}
  directed us to a custom table-based implementations
  the two mathematical functions.
Related comments, commented-out code, and older releases
  revealed that the developers of the 3.5 release, likely around 2004,
  concluded that the system implementations of these functions
  were too slow for their use case, and so wrote custom implementations
  to exploit two features of their use case:
  that POV-Ray only calls \F{sin} and \F{cos} with inputs between $-\pi$ and $\pi$;
  and that POV-Ray can tolerate significant inaccuracy in the result.
These custom implementations are used to compute the following expression,
  which is the input to \name:
\begin{equation}\label{povprog}
\begin{array}{l}
 \K{require}\: \theta, \phi \in [-\pi, \pi] \land n_x, n_y, n_z \in [-1, 1] \\
  \K{let}\: c = \cos(\theta), d_y = \sin(\theta)           \\
  \K{let}\:  d_x = c \cdot \cos(\phi), d_z = c \cdot \sin(\phi) \\
  \K{return}\:  n_x \cdot d_x + n_y \cdot d_y + n_z \cdot d_z
\end{array}
\end{equation}
This expression is the ``photon incidence computation'' in POV-Ray's ``caustics'' module.
Caustics are light effects like the lensing effects of glass or the
  pattern at the bottom of a swimming pool;
  to model these, POV-Ray shoots virtual photons at the scene
  and calculates how these photons move using the photon incidence computation.
The expression above computes the reflected energy
  of an incoming photon (with direction given by $\theta$ and $\phi$)
  reflected from a surface with normal vector $n$.
Millions of photon paths must be used to achieve realistic results,
  so photon incidence computation is a bottleneck,
  responsible for as much as \nPovRayPhotonsPercent of POV-Ray's total runtime
  when caustics are enabled.%
\footnote{
Speed is so important to POV-Ray
  that the caustics model FAQ includes advice
  on tuning the number of photons to trade off
  between quality and run time~\cite{povray-wiki}.
Many POV-Ray scene files don't even enable caustics,
  lowering the quality of the resulting render.
}

\subsection{Changing Function Implementations}

\begin{figure}
  \begin{subfigure}[t]{.48\linewidth}
    \includegraphics[width=\linewidth]{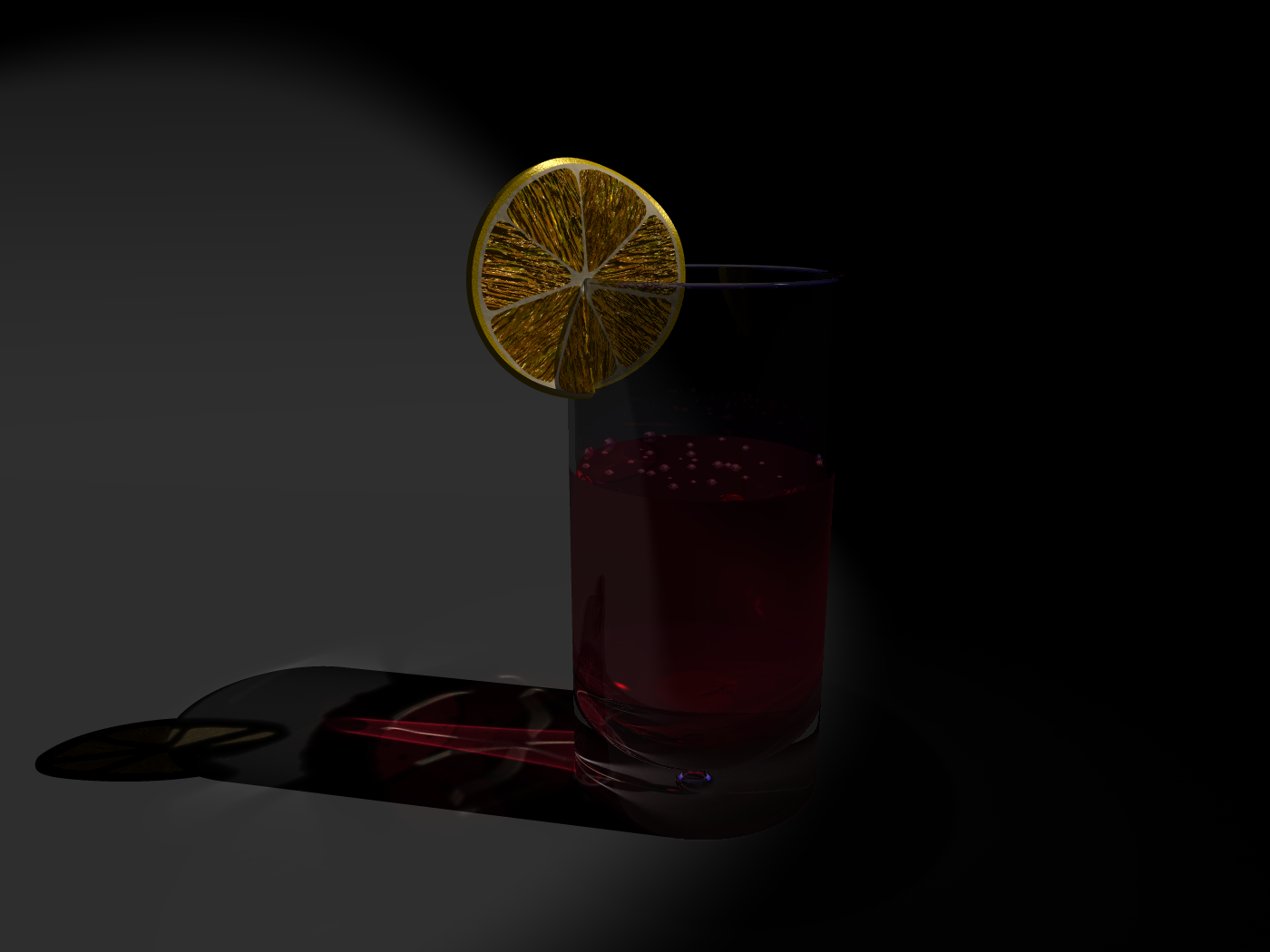}
    \caption{\F{grenadine} rendered using GLibC \F{sin} and \F{cos}.}
    \label{fig:grenadine-glibc}
  \end{subfigure}%
  \hfill%
  \begin{subfigure}[t]{.48\linewidth}
    \includegraphics[width=\linewidth]{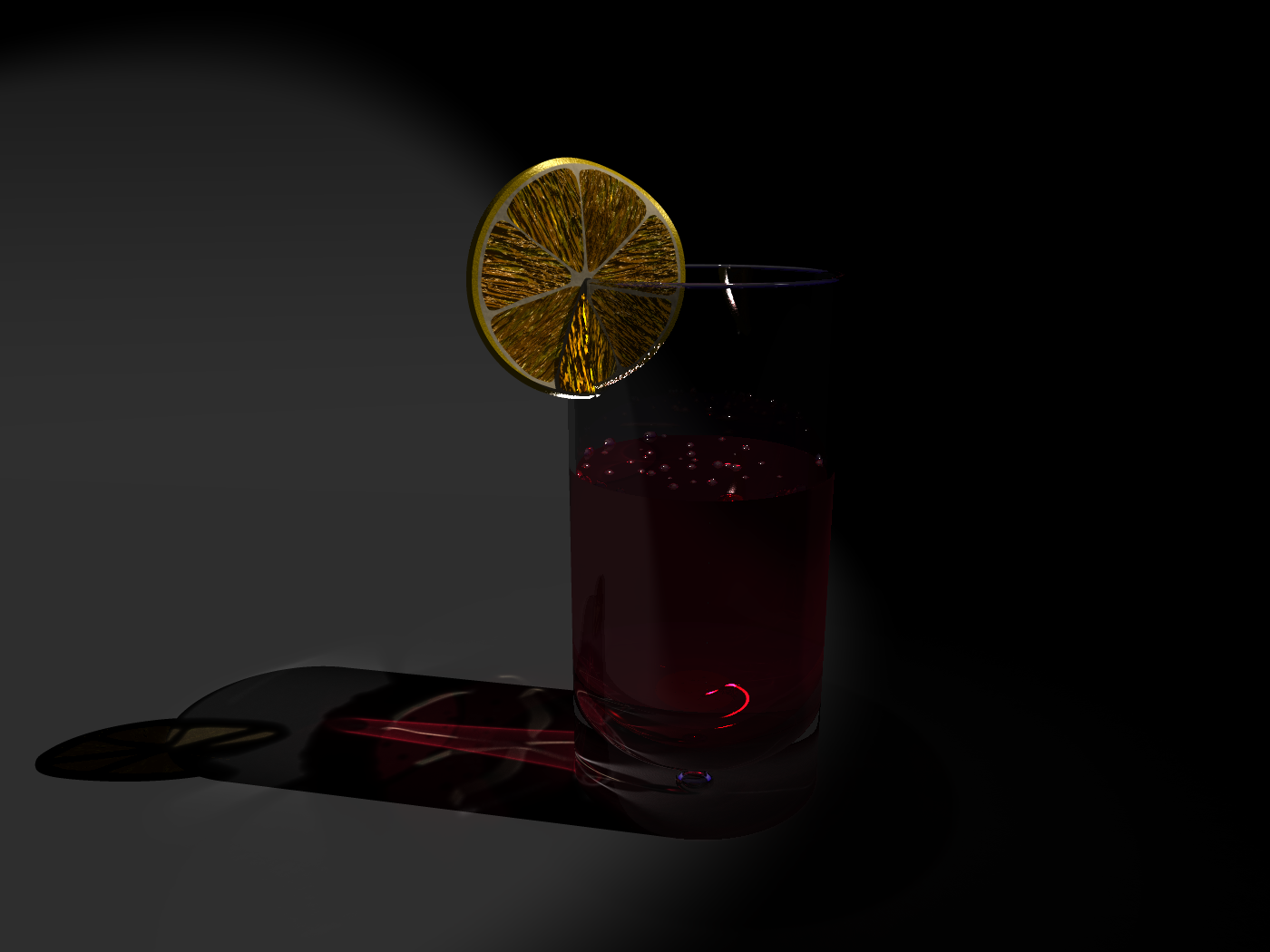}
    \caption{Same with $\F{sin}(x) = x$ and $\F{cos}(x) = -x$.}
    \label{fig:grenadine-crappy}
  \end{subfigure}%
  \caption{
    Two renderings of the \F{grenadine} scene using POV-Ray.
    On the left, the reference rendering
      uses GLibC's implementations of \F{sin} and \F{cos},
      which is accurate but quite slow.
    On the right, using crude approximations of \F{sin} and \F{cos}
      is much faster, but leads to obvious visual artifacts
      (look at the the orange slice, cocktail surface, and glass bottom).
    We tested several variations and $\F{cos}(x) = -x$
      was the fastest ``crude'' approximation.
  }
  \label{fig:grenadines}
\end{figure}

Developing a custom \F{sin} and \F{cos} implementation
  for use in just this computation was a sharp insight
  on the part of the POV-Ray developers.
Photon incidence spends almost all of its time
  inside the \F{sin} and \F{cos} functions,
  and the custom \F{sin} and \F{cos} implementations
  are \nPovProgTableSpeedup faster%
\footnote{
Timing measurements in this section refer specifically to the version of POV-Ray
  included in the SPEC 2017 benchmark suite,
  though both modern versions as well as versions going back to 2004
  contain code substantially similar to that discussed.}
  than the system GLibC libraries.
These custom implementations use 255-entry tables
  of \F{sin} and \F{cos} values between $-\pi$ and $\pi$,
  similar to that shown in \Cref{fig:table-based}.
Of course, due to the limited size of the tables,
  the custom \F{sin} and \F{cos} are also
  significantly less accurate than the system libraries;
  but the increased speed was worth it
  to the POV-Ray developers.
This optimization demonstrates the expertise
  that the POV-Ray developers are fortunate enough to poses.

But can we do better---can we find
  even faster and more accurate implementations of \F{sin} and \F{cos}
  for this particular expression?
For example, $\sin(x) \approx x$, at least for $x \approx 0$;
  implementing \F{sin} and \F{cos} with such crude approximations
  speeds up POV-Ray by another \nPovRayConstSpeedup
  over and above the POV-Ray developers' version.
But now the cost in accuracy is too steep:
  a standard test scene, \F{grenadine},
  rendered with this ultra-fast \F{sin} implementation
  looks like \Cref{fig:grenadine-crappy},
  with unrealistic highlights and lighting effects
  where glass interacts with the water in the drink or in the orange.
The resulting render is
  very different from the ground truth in \Cref{fig:grenadine-glibc}
  and looks worse than without caustics enabled at all.

Fortunately, there are dozens of implementations of \F{sin}
  between the extremes of GLibC and $\sin(x) \approx x$.
\name's linear \textit{error models} provide a simple way
  to quantify the effect of different function implementations
  and thereby search this space of possibilities.
\name computes \Cref{povprog}'s error model as:
\begin{equation}\label{errmodel}
  \mathcal{E} =
  1.41 \varepsilon_{c1}
  + \varepsilon_{s1}
  + \varepsilon_{c2}
  + \varepsilon_{s2}
  + 1.41 \delta_{c1}
  + \delta_{s1}
  + \delta_{c2}
  + \delta_{s2}
  + \float{2.09}{-15}
\end{equation}
Here, $\mathcal{E}$ is the overall error of \Cref{povprog}
  and the $\varepsilon$ and $\delta$ variables
  measure the accuracy of each call to \F{sin} and \F{cos}.
For any given choice of \F{sin} and \F{cos} implementations,
  \Cref{errmodel} estimates the impact
  of that choice on accuracy, which can be used
  to efficiently search for good implementation choices.
Note that unlike the error models in \Cref{sec:big-idea},
  this error model includes a constant term, \float{2.09}{-15}.
This constant term represents error due from operators
  that \name cannot tune, in this case addition and multiplication.

\begin{figure}
  \begin{subfigure}{.31\linewidth}
      \includegraphics[width=\linewidth]{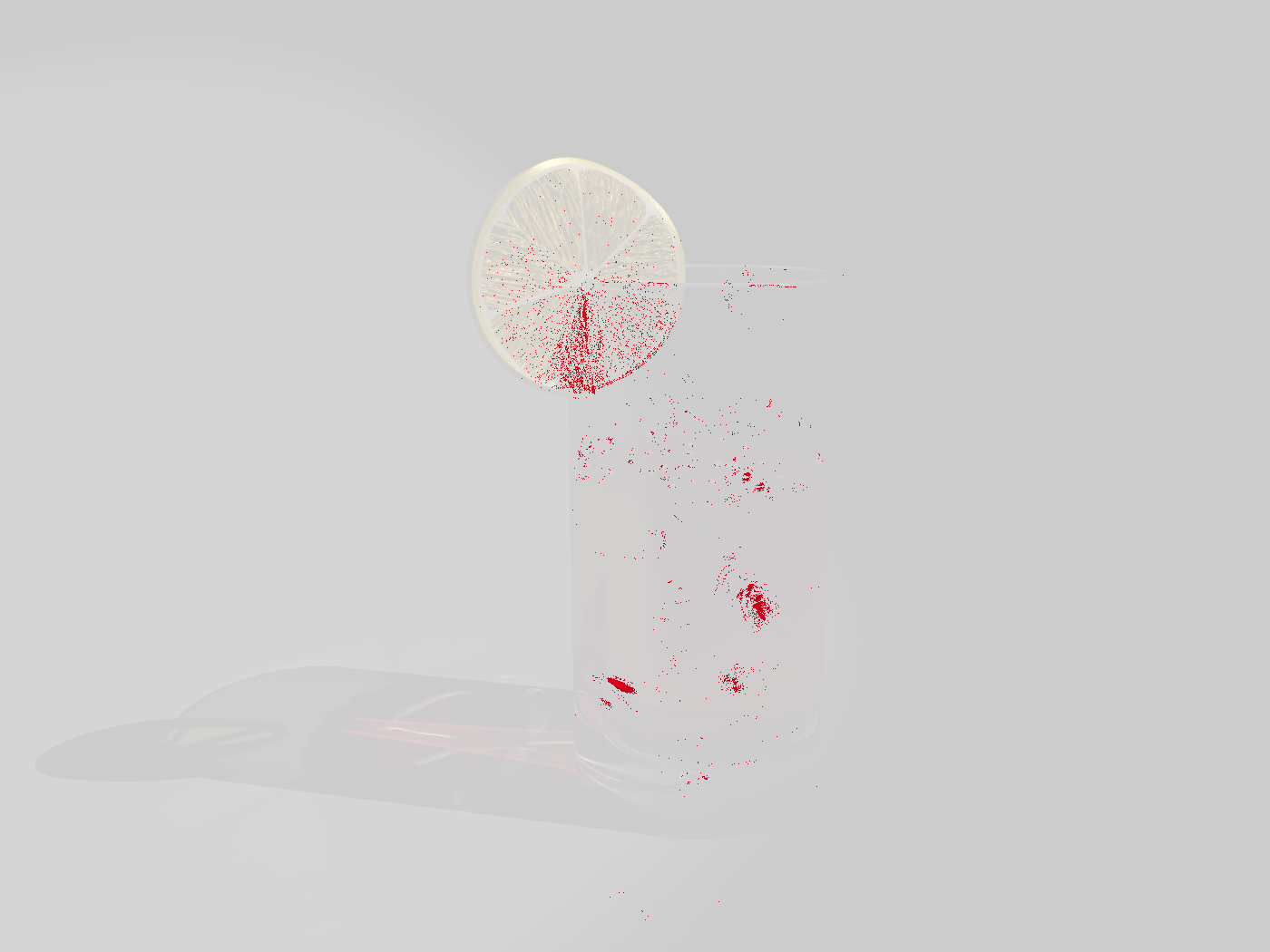}
      \caption{Errors from POV-Ray's table-based \F{sin} and \F{cos}.}
  \end{subfigure}%
  \hfill%
  \begin{subfigure}{.31\linewidth}
      \includegraphics[width=\linewidth]{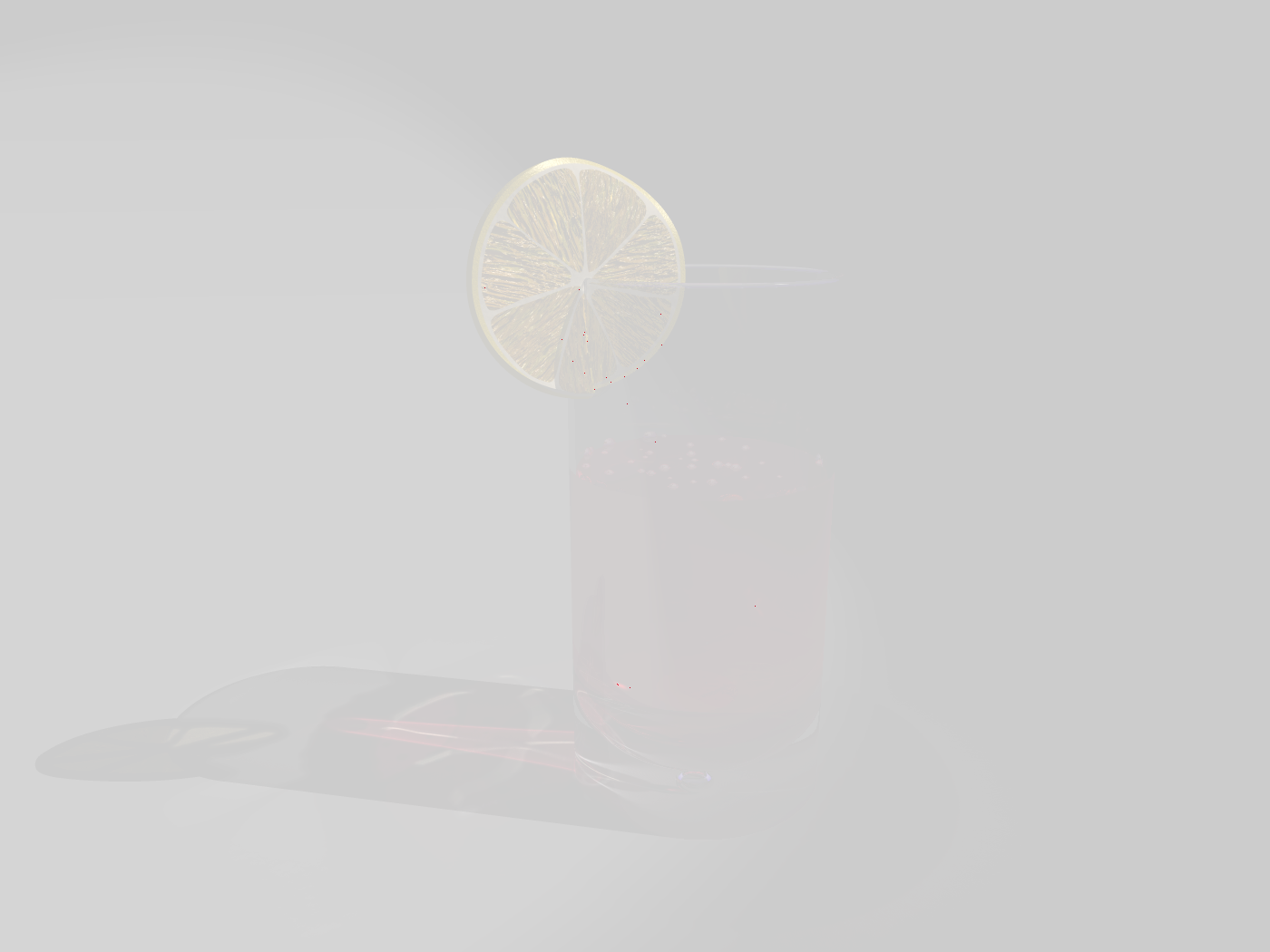}
      \caption{Errors from \name's suggested configuration.}
  \end{subfigure}%
  \hfill%
  \begin{minipage}{.31\linewidth}
  \caption{
      Differences from the reference rendering
        for two tuned versions of POV-Ray.
      On the left, POV-Ray's table-based \F{sin} and \F{cos}
        has significant, easily-visible errors.
      On the right, the chosen \name configuration
        produces minimal errors yet retains
        almost all of the table-based version's speed.
      \name makes finding such configurations easy.
  }
  \label{fig:grenadine-compare}
  \end{minipage}
\end{figure}

Plugging some values into the model gives a good feel
  for the scale of errors from using differing implementations.
For the double precision implementations of both \F{sin} an \F{cos} provided by
  GLibC the corresponding $\varepsilon$ and $\delta$ are
  $\float{2.22}{-16}$ and $\float{5}{-324}$.
Using these values gives an overall error of $\float{3.07}{-15}$,
  which goes some way toward explaining
  why using GLibC generates the correct render in \Cref{fig:grenadine-glibc}.
The POV-Ray developers' implementations
  have $\varepsilon = 0$ and $\delta = 0.02473$,
  leading to an overall error of $0.11$,
  $\float{3.56}{13}$ times larger than GLibC,
  explaining some of the errors seen in \Cref{fig:grenadine-compare}.
The crude approximations of $\sin(x) = x$ and $\cos(x) = -x$,
  meanwhile, produce an error of $16.28$
  (for a value actually between $-1$ and $1$),
  which explains the terrible rendering in \Cref{fig:grenadine-crappy}.
In other words, the error model makes it easy
  to estimate how choosing certain function implementations
  impacts the error of a floating-point expression.

\subsection{How \name Works}

The error model is convenient
  for sketching out the benefits of alternative function implementations.
Yet with two calls to \F{sin} and two calls to \F{cos},
  each of which could use a different implementation,
  \cref{povprog} has millions of possible configurations.
What's needed is a tool that uses \cref{errmodel}
  to automatically elevate implementation choices
  that optimally trade speed for accuracy.

\name does just this.
Since only the application developers can decide
  how much accuracy to trade for speed,
  \name outputs a \textit{speed-accuracy Pareto curve},
  where each point on the curve
  is the most accurate configuration possible at a given speed.
To derive this curve,
  \name combines the error model above
  with a simple profiled \textit{cost model}
  that estimates the time to evaluate each function implementation.
Importantly, both the error and cost models are linear,
  which allows \name to phrase the choice of implementation
  as an integer linear program.
\name can then use an off-the-shelf ILP solver
  to compute points along the speed-accuracy Pareto curve.
For example, \name can tune POV-Ray's photon incidence computation
  against a collection of \nSinImpls \F{sin} implementations
  and \nCosImpls \F{cos} implementations,
  a space of \nPovRayPossible possible configurations.
Out of this vast search space,
  \name produces the speed-accuracy Pareto curve
  shown in \Cref{fig:povray-expression},
  with just \nPovProgConfigs configurations,
  in \nPovProgOptunerTime.

Automating implementation selection with \name
  also makes novel optimizations possible.
For example,
  a function implementations that is only called
  on a restricted range of inputs
  can often be faster without being less accurate.%
\footnote{Handling large input ranges
  usually uses higher-precision arithmetic,
  such as with Cody-Waite range reduction~\cite{cody-waite},
  which is complex and expensive.}
\name automates this optimization
  by extending the integer linear program
  to also compute bounds on the value of any expression,
  allowing it to use restricted-range implementations
  when their arguments are in range.
To make use of this,
  \name includes custom, reduced-range implementations
  of \F{exp}, \F{log}, \F{sin}, \F{cos}, and \F{tan}
  that are much faster than traditional, full range, implementations.
Normally, reduced-range implementations are dangerous,
  since using them on out-of-range inputs can lead to disasterous results.
But the sound error bounds computed by \name
  make this complex and dangerous optimization easy and safe.

\subsection{Results}

\begin{figure}
  \begin{subfigure}{0.48\linewidth}
    \includegraphics[width=\linewidth]{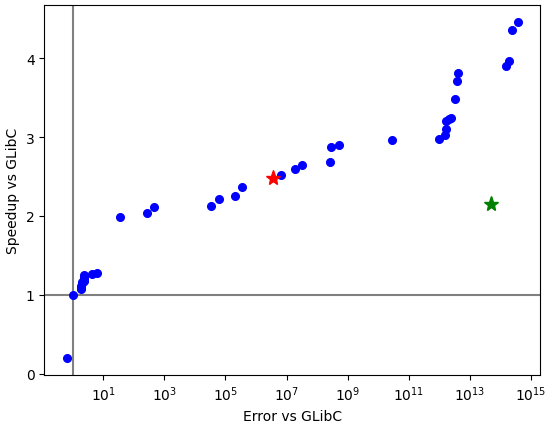}
    \caption{Speed-accuracy Pareto curve for \Cref{povprog}}
    \label{fig:povray-expression}
  \end{subfigure}%
  \hfill%
  \begin{subfigure}{0.48\linewidth}
    \includegraphics[width=\linewidth]{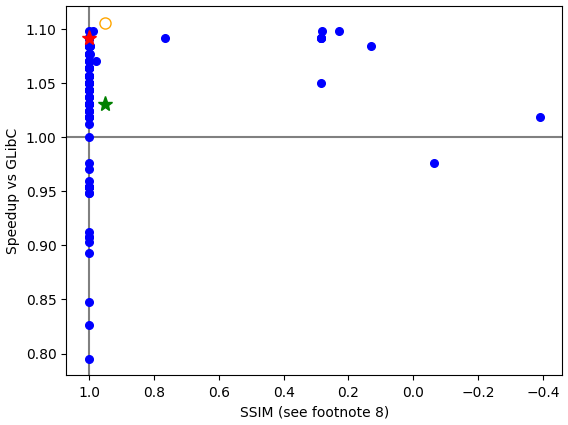}
    \caption{Speed-accuracy Pareto curve for POV-Ray}
    \label{fig:povray-end2end}
  \end{subfigure}
  \caption{Speed vs. accuracy for various configurations of \Cref{povprog}.
    On the left, speed and accuracy for just \Cref{povprog};
      on the right, speed and accuracy for POV-Ray as a whole
      (on the \F{grenadine} scene).
    Speedup is normalized to the all-GLibC configuration.
    \name's configurations are marked in blue and red,
      while the POV-Ray developers' custom implementation
      is marked with a green star.
    \name's tuned configurations are both faster and more accurate
      than the developers' custom implementation.
    The orange circle in the graph on the right
      relates to future work described in \Cref{sec:cs-methods}.
    }
\end{figure}

The \nPovProgConfigs configurations chosen by \name
  are shown in \Cref{fig:povray-expression}.
Each point in that plot is a configuration---%
  a choice of library implementation for each call to \F{sin} and \F{cos}---%
  and its location on the plot gives that configuration's speed
  and worst-case accuracy bound.
These configurations range from
  an extreme-accuracy configuration,
  which uses the verified CRLibM library for each call site
  and is \nPovProgCrlibmAccuracyIncrease more accurate
  and \nPovProgCrlibmSlowdown slower than GLibC,
  to an extreme-speed configuration,
  which uses custom implementations
  and is \nPovProgFastSpeedup faster and
  \nPovProgFastAccuracyDecrease less accurate than GLibC
  (that is, it has a relative error of \nPovProgFastRelativeError).
Somewhere in between these extremes
  is the red starred point in \Cref{fig:povray-expression},
  which uses custom order 13/15 implementations for the \F{sin} calls
  and a custom order 14 implementation for \F{cos},
  all fit to the $[-\pi, \pi]$ input range.
This configuration is \nPovProgLosslessSpeedup faster
  and \nPovProgLosslessAccuracyDecrease less accurate
  than the GLibC configuration;
  that makes it both faster and more accurate
  than the POV-Ray developers' custom table-based \F{sin} and \F{cos},
  shown with a green star in \Cref{fig:povray-expression}.

Using \name's suggested configurations
  can lead to end-to-end application speedups.
We produced modified versions of POV-Ray
  for each of \name's suggested configurations
  and plotted both their overall run time
  and the quality of their outputs in \Cref{fig:povray-end2end}.%
\footnote{
  Image quality is measured using
  the structural similarity index measure~\cite{ssim},
  a standard measure of image quality.
  In particular, we use the SPEC 2017 benchmarking harness
  to measure both SSIM and runtime.
  More details of the methodology are given
  in \Cref{sec:cs-methods}
}
POV-Ray is a naturally-error-tolerant program
  as shown by the blue streak on the left-hand-side of the figure:
  \name's \nPovRaySatConfigs most accurate configurations
  all render the reference image exactly.
Likewise, there's a limit to how much POV-Ray can be sped up
  just by changing function implementation,
  for Amdahl's-law-like reasons.
The red starred configuration hits both these limits,
  with results identical to the reference rendering,
  but produced roughly \nPovRayLosslessSpeedup faster.
By contrast, the green starred point,
  which is the POV-Ray developers' implementation,
  is both slower and has noticable errors
  (\Cref{fig:grenadine-compare}).
Of course, it is up to the POV-Ray developers to decide
  how much accuracy loss is acceptable
  and how large a speed-up makes up for a given level of error.
But here, \name's proposed configurations
  are simultaneously faster and more accurate.
Moreover, the point marked in red is not the only option
  produced by \name;
  the developers can experiment with different configurations
  and easily find one that is faster and less accurate
  if they so desire.

\section{Mathematical Function Implementation Background}
\label{sec:implementations}

While the mathematics behind approximating transcendental functions
  is well understood,
  numerous choices, like approximation method,
  polynomial order, table size, and range reduction strategy,
  all impact both accuracy and speed.

\begin{figure}
\begin{subfigure}{0.60\textwidth}
\includegraphics[width=\textwidth]{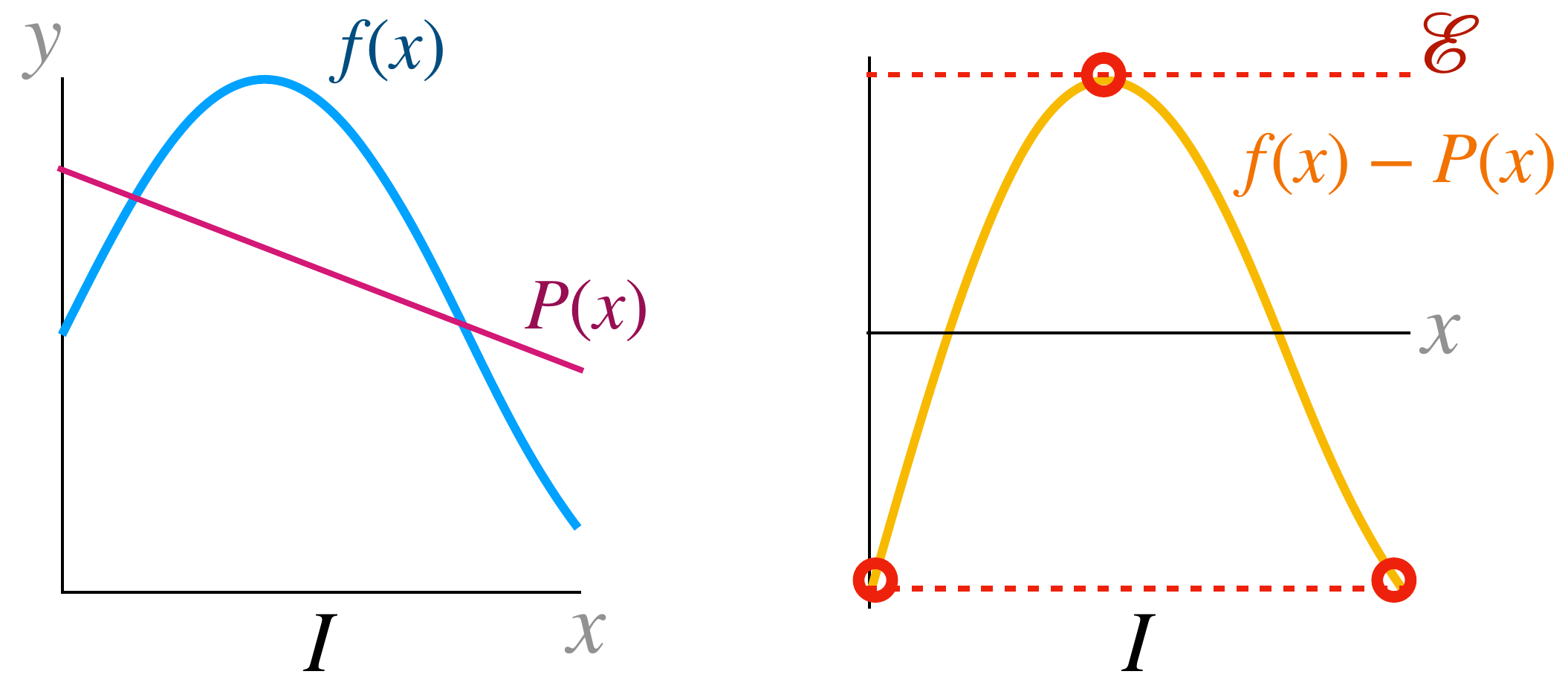}
\caption{Polynomial approximation and error}
\label{fig:poly}
\end{subfigure} \hfill%
\begin{subfigure}{0.315\textwidth}
\includegraphics[width=\textwidth]{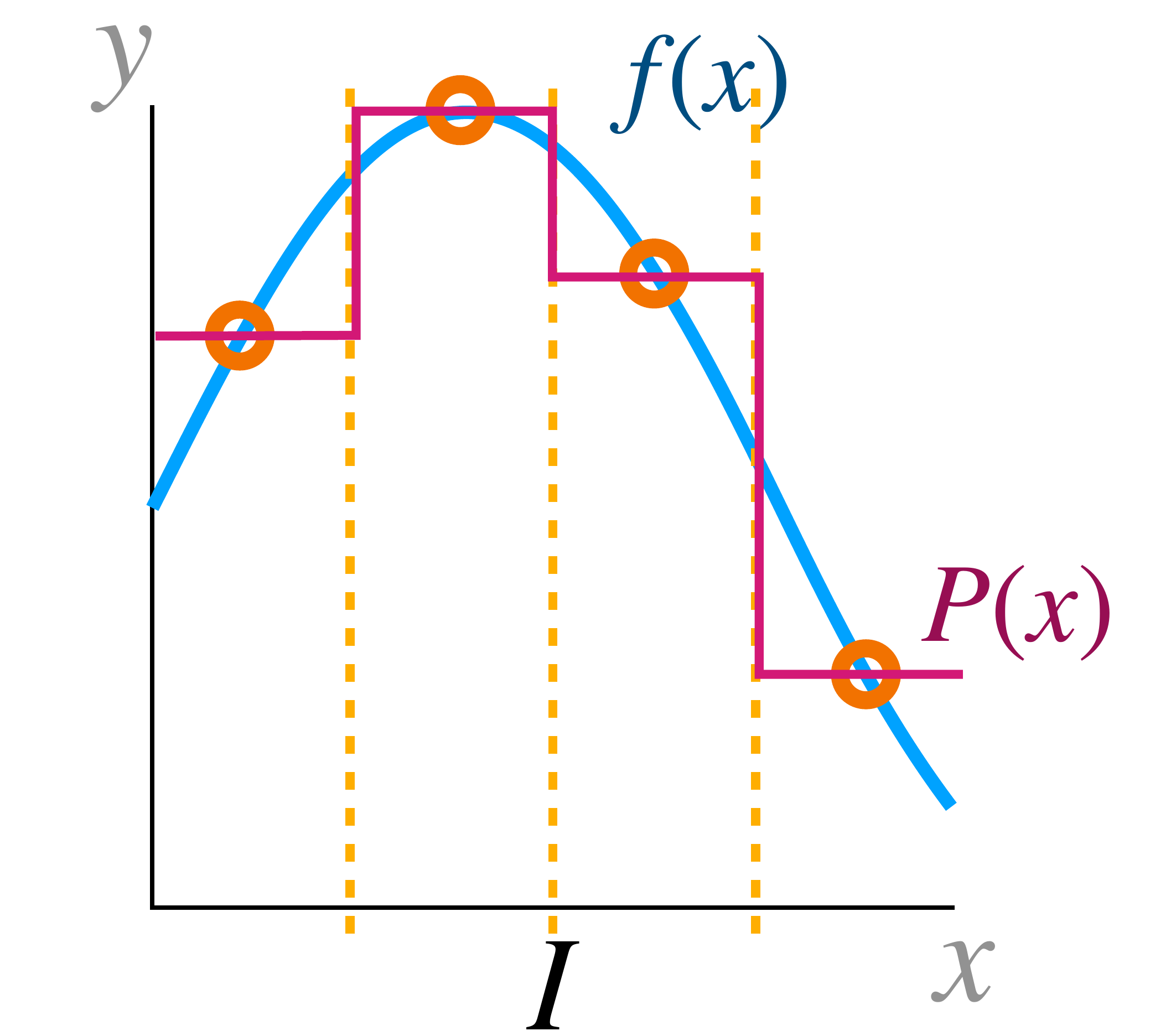}
\caption{Table approximation}
\label{fig:table-approx}
\end{subfigure}
\caption{
Approximations $P(x)$ of an arbitrary function $f(x)$.
A order-1 polynomial approximation is shown on the left;
  a $4$-entry table based approximation is shown on the right.
In the middle, an illustration of the equioscillation theorem:
  the polynomial approximation is optimal
  error reaches its highest value the maximum
  number of times.
}
\label{fig:approx}
\end{figure}

\begin{figure}
\begin{subfigure}{0.48\textwidth}
\scriptsize
\begin{verbatim}
double ml2_raw_wide_sin_13(double x){
  double x2 = x * x;
  double pa, pa1, pa3, pa5, pa7, pa9, pa11, pa13;
  pa13 =  0x1.52a851954275cp-33;
  pa11 = -0x1.ae00bdd2a86a8p-26 + x2 * pa13;
  pa9  =  0x1.71dce463cf737p-19 + x2 * pa11;
  pa7  = -0x1.a019fce360596p-13 + x2 * pa9;
  pa5  =  0x1.11111109020a6p-7  + x2 * pa7;
  pa3  = -0x1.5555555540916p-3  + x2 * pa5;
  pa1  =  0x1.ffffffffffdc9p-1  + x2 * pa3;
  pa   = x * pa1;
  return pa;
}
\end{verbatim}
\caption{An order-13 polynomial function approximation}
\label{fig:poly-based}
\end{subfigure}%
\hfill%
\begin{subfigure}{0.48\textwidth}
\scriptsize
\begin{verbatim}
double sin_table[255];

void fill_table() {
  for(i=0; i<255; i++){
    theta = (double)(i-127)*M_PI/127.0;
    sin_table[i] = sin(theta);
  }
}

double table_sin(double x){
  int index = (x*127.0/M_PI) + 127;
  return sin_table[index];
}
\end{verbatim}
\caption{A size-255 table-based function approximation}
\label{fig:table-based}
\end{subfigure}
\caption{
  Two possible approximations of \F{sin} for $x \in [-\pi, \pi]$.
  The one on the left uses an order-13 order polynomial
    and achieves an absolute error of \nScarySinError
    and a run time of \nScarySinTime on our reference machine.
  The on on the right
    is the table based implementation used by POV-Ray,
    with an absolute error of $0.02473$
    and a run time of \nPovRayTableImplTime.}
\end{figure}

\subsection{Approximating over an Interval}

The core technique in implementing a mathematical function
  is approximating a function $f$ over a small input range $I$.
One method is to find a polynomial $P(x)$ that approximates $f$.
While techniques like Taylor series and Chebyshev approximations,
  which optimize for the average case, are better known,
  implementations usually optimize for worst-case errors.
The Remez exchange algorithm~\cite{remez},
  which solves for local minima and maxima of $f(x) - P(x)$
  by the Chebyshev equioscillation theorem~\cite{equioscillation},
  is the standard way of deriving coefficients for $P$
  that minimize worst-case error.
\Cref{fig:poly} illustrates this approach.
Remez exchange is implemented in the the Sollya tool~\cite{sollya},
  which additionally uses the LLL algorithm~\cite{lll}
  to tune the coefficients for the floating-point domain.
The end result of this process is an implementation
  similar to the one shown in \Cref{fig:poly-based}.
The number of terms in the polynomial is the key parameter to this process;
  an implementation can approximate a target functions like $\sin$ or $\exp$
  with anywhere from one to dozens of terms,
  offering a spectrum with higher-order approximations
  being more accurate but also slower.
The input interval $I$ is also an important input;
  generally speaking, wider input intervals yield worse approximations.
Even once a polynomial is found, there are more choices to make.
For example, the polynomial
\(
  P(x) = a_0 + a_1 \cdot x + a_2 \cdot x^2 + \dotsb
\)
is usually evaluated via Horner's rule,
\(
  P(x) = a_0 + x\cdot(a_1 + x\cdot(a_2 + x\cdot(\dotsb)))
\),
  but there are other evaluation schemes as well,
  like Estrin’s Method for parallelization with SIMD
  or compensated summation for higher accuracy.

Table-based implementations are an alternative to pure polynomial approximation.
In a table-based implementation,
  the interval $I$ is split into many smaller intervals,
  often hundreds or thousands of them,
  and then some uniform method is used to approximate the function
  on each of those smaller intervals.
A simple table-based implementation like this
  tabulates $f(x)$ for evenly spaced $x$ values,
  like the $\sin$ table in the back of an old-school math textbook;
  such an implementation is shown in \Cref{fig:table-based}.
A more sophisticated approach
  may store polynomial coefficients in the table,
  or use complex interpolation schemes to fill in intermediate values.
Just as with polynomial implementations,
  a table-based scheme has many parameters for the implementor to choose:
  table size, interpolation scheme, subdivision algorithm, and so on;
  and again, these parameters affect both speed and accuracy,
  and may depend on machine-specific parameters like cache size.

\subsection{Range Reduction}

Neither polynomial nor table-based approximation
  works well over a large input interval $I$,
  so the most challenging part of approximating a mathematical function
  is making sure the function can be applied to an arbitrary input.
This is called \textit{range reduction and reconstruction}.
Range reduction and reconstruction bookend the polynomial or table,
  transforming the input to lie within the polynomial or table's input range
  and then adjusting the output to fit the original input.

\begin{figure}
\hfill%
\begin{minipage}[c]{.56\textwidth}
\includegraphics[width=\textwidth]{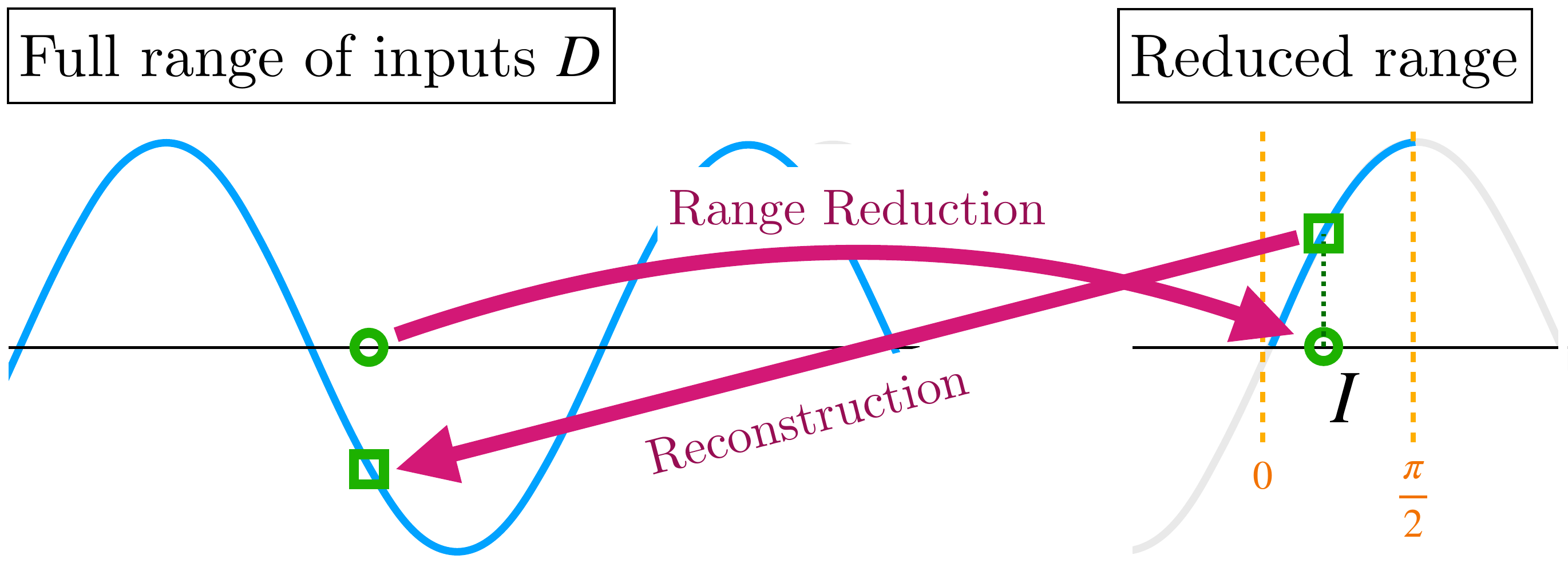}
\end{minipage}\hfill
\begin{minipage}[c]{0.38\textwidth}
\caption{
\small
Range reduction and reconstruction for \F{sin}.
The input point, somewhere along the input domain, is mapped via range reduction
  into the interval $[0,\pi/2]$ for evaluation, then mapped back via
  reconstruction to the correct output value.
This reduction takes advantage of the cyclic nature of \F{sin}, the symmetry of
  the sections of \F{sin} above and below $y=0$, and the mirror symmetry
  present in each half cycle.
Reconstruction of \F{sin} needs only determine the correct sign of the output.
}
\end{minipage}\hfill
\label{fig:range}
\end{figure}

Consider implementing \F{sin} over a large range.
Since \F{sin} is periodic, $\sin(x) = \sin(x \bmod 2\pi)$;
  and furthermore, $x \bmod 2\pi$ is bounded
  to the range $[0, 2\pi]$.
Thus, \F{sin} can be implemented over a large range
  by using a polynomial or table for the range $[0, 2\pi]$
  and \emph{reducing} other inputs to lie within that range.
We can take advantage of other symmetries
  to restrict the range further.
For example, $\sin(x) = -\sin(x - \pi)$;
  for inputs $x \in [\pi, 2\pi]$ the right hand side of this equation
  calls $\sin$ on inputs $x - \pi \in [0, \pi]$.
So, a polynomial or table fit to $[0, \pi]$ is enough,
  but in this case a \emph{reconstruction} step is necessary
  to take the output on the reduced input
  (in this case, $\sin(x - \pi)$)
  and compute from it the output for the original input
  (in this case, by negating the reduced output).
Both reduction and reconstruction are shown in \cref{fig:range}.
Other identities, like the double-angle formula,
  can also be used for range reduction,
  and the same general principle
  of using function identities to reduce the input range
  can apply to $\log$, $\exp$, or any other functions
  with identities to leverage.

Reducing the interval $I$ that the polynomial or table is fit to
  usually makes the approximation itself
  either more accurate or faster (by using fewer terms).
But range reduction and  reconstruction
  can slow the implementation down
  and add error of their own.
Traditionally, math libraries have used higher-precision arithmetic,
  such as in Cody \& Waite reduction~\cite{cody-waite},
  to implement range reduction with no error.
This maximizes accuracy, especially for very large inputs
  (imagine computing $\F{sin}(10^{10})$),
  but is also very slow because it requires computing $x \bmod 2\pi$
  in very high precision using thousands of bits of $\pi$.
But range reduction can also trade accuracy for speed.
For example, $x \bmod 2\pi$ can be computed
  by just evaluating $x - \lfloor x / 2 \pi \rfloor$ in double precision.
This introduces errors
  (from approximating $\pi$ and from division)
  but is significantly faster than high-precision arithmetic.
There are also intermediate choices with more or less error.
Thus, the choice of range reduction
  again presents a trade-off between accuracy and speed.

\section{Error and cost models}
\label{sec:error-and-cost-models}

The input to \name is a floating-point expression $\tilde{E}$,
  with no loops or control flow,
  over some number of floating-point inputs;
  for simplicity, this section refers to a single input $x_0$,
  though generalizing to multiple inputs is straightforward.
In addition to the expression $\tilde{E}$,
  the input must also contain an interval $I$
  that the input $x_0$ is drawn from.
The core idea behind \name,
  as described in \Cref{sec:big-idea},
  is the linear error model for a floating-point expression.
To compute the error model \name uses error Taylor series~\cite{fptaylor}.

\subsection{Error model}

Take a floating-point expression $\tilde{E}$
  generated by the grammar $e ::= \tilde{f}(e, e, \ldots) \mid x_0$.
In this paper, variables with tildes, like $\tilde{x}_i$,
  represent floating-point values,
  and the same variables without tildes, like $x_i$,
  represent their true mathematical value
  when computed without rounding error.
Each $\tilde{f}$ thus implements of some underlying function $f$;
  floating-point constants like \F{PI}
  can be considered zero-argument functions.

Transform the expression $\tilde{E}$ to a linear sequence of function calls
  $\tilde{x}_n := \tilde{f_n}(\tilde{x}_{n-a_n}, \tilde{x}_{n-b_n}, \ldots)$,
  where later function calls
  may use the results of earlier calls as arguments,
  and where the value of the expression as a whole
  is the value of the last variable in the sequence, $\tilde{x}_N$.
By \cref{rdmodel},
  $\tilde{x}_n = f_n(\tilde{x}_{n-a_n}, \tilde{x}_{n-b_n}, \ldots)(1 + \varepsilon_n) + \delta_n$.
Note that $\tilde{x}_N$ is thus a function of the $\varepsilon_n$s and $\delta_n$s.
When all $\varepsilon_n = \delta_n = 0$,
  the result is $x_n$ computed without any error;
  but $\varepsilon_n$ and $\delta_n$ are not equal to zero.
Instead they are small, unkown values
  bounded by constants $\varepsilon_{f_n}$ and $\delta_{f_n}$.

Define the constants
\[
A^k_n = \max_{x \in I} \left( \frac{\partial \tilde{x}_k}{\partial \varepsilon_n}
\left[\begin{array}{l}0 / \varepsilon \\ 0 / \delta \end{array}\right] \right)
\;\text{and}\;
B^k_n = \max_{x \in I} \left( \frac{\partial \tilde{x}_k}{\partial \delta_n}
\left[\begin{array}{l}0 / \varepsilon \\ 0 / \delta \end{array}\right] \right)
\]
  and $A_n = A^N_n$ and $B_n = B^N_n$.
Note that these constants are real numbers
  and can be computed
  by a one-dimensional global nonlinear optimization;
  or, for expressions with multiple variables,
  over as many dimensions as there are input variables.
Then the error of $\tilde{E}$ is:
\begin{equation}\label{errmodelformal}
  |\tilde{x}_N - x_N| \le \sum_n A_n \varepsilon_{f_n} + \sum_n B_n
  \delta_{f_n} + o(\varepsilon^2).
\end{equation}
Here, the $o(\varepsilon^2)$ term represents
  the quadratic and higher-order terms
  of the error Taylor series.
It can be bounded using Lagrange's theorem,
\[
  o(\varepsilon^2) \le
  \max_{x \in I} \left(
  \sum_{u, v \in \{\varepsilon, \delta\}}
  \sum_{i,j}
  \frac{\partial^2 x_k}{\partial u_i \partial v_i}
    \left[\begin{array}{l}0 / \varepsilon \\ 0 / \delta \end{array}\right]
    u_{f_i}
    v_{f_j}
    \right),
\]
  where the sums range over all pairs of $\varepsilon/\delta$ variables.
Importantly, since this bound involves higher powers
  of the $\varepsilon_f$s and $\delta_f$s,
  it tends to be insignificantly small.
It is necessary for soundness,
  but is not particularly important for estimating the error.

\Cref{errmodelformal},
  including the Lagrange bound for the higher-order terms,
  is the traditional use of error Taylor series.
\name instead drops the higher-order terms
  and considers the remaining linear terms as
  a function of the $\varepsilon$ and $\delta$ variables
  that estimates $\tilde{E}$'s error
  in terms of the function implementations it uses.
Note that typically, an expression contains operators
  like addition and multiplication that \name cannot tune.
In this case
  some of the $\varepsilon$ and $\delta$s are fixed
  and \name folds those resulting terms into a constant.

\subsection{Cost model}

The complement to the linear error model is a linear cost model,
  which estimates the speed of the expression
  given the implementations chosen for each function call.
\name uses a simple cost model:
\[
\mathit{cost} = \sum_n c_{f_n},
\]
  where $c_f$ is the average runtime of the function implementation $\tilde{f}$.
To measure $c_f$, each implementation
  is compiled and run
  on \nTimingInputs random valid inputs
  in a tight loop for \nTimingElapsed.
To try to get maximally accurate timings.
  we use multiple measurements, input pre-generation, core binding,
  and the \texttt{-O3 -march=native -mtune=native -DNDEBUG} compiler flags.

Normally, a simple sum of average runtimes is too simplistic:
  it ignores the complexities of modern CPUs
  and the details of input-dependent control flow.
Several factors make this model more appropriate in our setting.
\name's goal is to find
  all Pareto-optimal trade-offs between the error and cost models,
  so the cost model only needs to be a \textit{relative order},
  not a precise prediction of runtimes.
Plus, mathematical function implementations usually
  have rare, easily-branch-predicted control flow;
  use few if any data structures or complex memory access patterns;
  and are generally compute-bound.
That makes average runtimes more meaningful
  than for general-purpose code.
Finally, since \name only changes the implementation of mathematical functions,
  any computation outside a function implementation
  will be the same across all configurations.
The cost model thus does not need to model those costs,
  and is only predicting the costs of calls to shared libraries.

One caveat is necessary for achieving good results with this cost model:
  the linear sequence of function calls $x_n = \tilde{f}_n(x_{n-a_n}, x_{n-b_n}, \ldots)$
  must contain no duplicate entries representing common subexpressions.
Compilers commonly perform common subexpression elimination
  to avoid recomputing the same expression twice.
Thus, if two subexpressions both compute $f(x_1, x_2)$,
  and the first subexpression uses implementation $\tilde{f}_a$
  while the second uses implementation $\tilde{f}_b$,
  the cost is $c_a + c_b$ if these implementations are different
  but only $c_a$ is both implementations are the more accurate of the two.
In other words, when common subexpressions exist,
  it's best to use the same, more accurate implementation at both sites.
\name thus performs common subexpression elimination
  when converting expressions to a linear sequence of function calls,
  and the cost and error models are built on this deduplicated sequence,
  ensuring that common subexpressions use the same implementations.

\subsection{Mathematical Libraries}
\label{ssec:libraries}

\newcommand{\mr}[2]{\multirow{#1}{*}{#2}}
\newcommand{\mrd}[2]{\multirow{#1}{2em}{#2}}

\begin{figure}
\begin{subfigure}[t]{0.5\linewidth}
{\footnotesize
\begin{tabular}[t]{|c|l|r|r|l|} \hline
Func         & Domain                                   & Error       & Cost  & Library  \\ \hline
\mr{5}{exp}  & \mrd{5}{$[-1.79e^{308},$ $709.78]$}      & 0.5         & 54.02 & CRLibM   \\ \cline{3-5}
             &                                          & \mr{3}{1.0} & 10.71 & GLibC    \\ \cline{4-5}
             &                                          &             & 10.70 & OpenLibm \\ \cline{4-5}
             &                                          &             & 10.78 & AMD LibM \\ \cline{3-5}
             &                                          & 5.0         & 5.32  & VDT      \\ \hline
\mr{5}{log}  & \mrd{5}{$[5.0e^{-324},$ $1.79e^{30}]$}   & 0.5         & 32.93 & CRLibM   \\ \cline{3-5}
             &                                          & \mr{3}{1.0} & 8.53  & GLibC    \\ \cline{4-5}
             &                                          &             & 8.53  & OpenLibm \\ \cline{4-5}
             &                                          &             & 8.36  & AMD LibM \\ \cline{3-5}
             &                                          & 5.0         & 5.99  & VDT      \\ \hline
\mr{6}{sin}  & \mrd{5}{$[-1.79e^{308},$ $1.79e^{308}]$} & 0.5         & 35.27 & CRLibM   \\ \cline{3-5}
             &                                          & \mr{3}{1.0} & 8.76  & GLibC    \\ \cline{4-5}
             &                                          &             & 8.76  & OpenLibm \\ \cline{4-5}
             &                                          &             & 7.56  & AMD LibM \\ \cline{3-5}
             &                                          & 5.0         & 4.42  & VDT      \\ \cline{2-5}
             & \mr{1}{\tiny$[-0.78, 0.78]$}             & 5.0         & 1.82  & VDT      \\ \hline
\mr{6}{cos}  & \mrd{5}{$[-1.79e^{308},$ $1.79e^{308}]$} & 0.5         & 34.35 & CRLibM   \\ \cline{3-5}
             &                                          & \mr{3}{1.0} & 8.84  & GLibC    \\ \cline{4-5}
             &                                          &             & 8.82  & OpenLibm \\ \cline{4-5}
             &                                          &             & 7.19  & AMD LibM \\ \cline{3-5}
             &                                          & 5.0         & 4.20  & VDT      \\ \cline{2-5}
             & \mr{1}{\tiny$[-0.78, 0.78]$}             & 5.0         & 2.04  & VDT      \\ \hline
\mr{5}{tan}  & \mrd{5}{$[-1.79e^{308},$ $1.79e^{308}]$} & 0.5         & 80.20 & CRLibM   \\ \cline{3-5}
             &                                          & \mr{3}{1.0} & 15.81 & GLibC    \\ \cline{4-5}
             &                                          &             & 16.00 & OpenLibm \\ \cline{4-5}
             &                                          &             & 28.66 & AMD LibM \\ \cline{3-5}
             &                                          & 5.0         & 8.90  & VDT      \\ \hline
\end{tabular}
}
\subcaption{Double precision implementations}
\end{subfigure}\hfill%
\begin{subfigure}[t]{0.5\linewidth}
{\footnotesize
\begin{tabular}[t]{|c|l|r|r|l|} \hline
Func         & Domain                                 & Error                       & Cost & Library  \\ \hline
\mr{5}{expf} & \mrd{5}{$[-3.40e^{38},$ $88.72]$}      & $\float{2.68}{8}$           & 5.06 & RLibM    \\ \cline{3-5}
             &                                        & \mr{3}{$\float{4.03}{8}$}   & 8.46 & GLibC    \\ \cline{4-5}
             &                                        &                             & 4.86 & OpenLibm \\ \cline{4-5}
             &                                        &                             & 4.68 & AMD LibM \\ \cline{3-5}
             &                                        & $\float{1.76}{10}$          & 5.20 & VDT      \\ \hline
\mr{5}{logf} & \mrd{5}{$[1.40e^{-45},$ $3.40e^{38}]$} & $\float{2.68}{8}$           & 6.26 & RLibM    \\ \cline{3-5}
             &                                        & \mr{3}{$\float{2.76}{8}$}   & 7.15 & GLibC    \\ \cline{4-5}
             &                                        &                             & 7.13 & OpenLibm \\ \cline{4-5}
             &                                        &                             & 5.88 & AMD LibM \\ \cline{3-5}
             &                                        & $\float{5.45}{8}$           & 5.80 & VDT      \\ \hline
\mr{6}{sinf} & \mrd{5}{$[-3.40e^{38},$ $3.40e^{38}]$} & \mr{3}{$\float{2.16}{9}$}   & 7.13 & GLibC    \\ \cline{4-5}
             &                                        &                             & 7.10 & OpenLibm \\ \cline{4-5}
             &                                        &                             & 7.64 & AMD LibM \\ \cline{3-5}
             &                                        & $\float{1.74}{10}$          & 4.69 & VDT      \\ \cline{2-5}
             & \mr{1}{\tiny $[-0.785, 0.785]$}        & $\float{1.74}{10}$          & 1.60 & VDT      \\ \hline
\mr{6}{cosf} & \mrd{5}{$[-3.40e^{38},$ $3.40e^{38}]$} & \mr{3}{$\float{2.16}{9}$}   & 7.13 & GLibC    \\ \cline{4-5}
             &                                        &                             & 7.10 & OpenLibm \\ \cline{4-5}
             &                                        &                             & 6.76 & AMD LibM \\ \cline{3-5}
             &                                        & $\float{1.74}{10}$          & 4.47 & VDT      \\ \cline{2-5}
             & \mr{1}{\tiny $[-0.785, 0.785]$}        & $\float{1.74}{10}$          & 1.93 & VDT      \\ \hline
\mr{5}{tanf} & \mrd{5}{$[-3.40e^{38},$ $3.40e^{38}]$} & \mr{3}{$\float{5.37}{8}$}   & 8.50 & GLibC    \\ \cline{4-5}
             &                                        &                             & 8.49 & OpenLibm \\ \cline{4-5}
             &                                        &                             & 7.03 & AMD Libm \\ \cline{3-5}
             &                                        & $\float{1.77}{10}$          & 8.32 & VDT      \\ \hline
\end{tabular}
}
\subcaption{Single precision implementations}
\end{subfigure}

\caption{
  The standard library implementations available to \name.
  Errors were found via documentation and source code comments
    and are given in ``units in the last place'';
    the actual $\varepsilon_f$ and $\delta_f$ values
    are the listed error times $2^{-52}$ for $\varepsilon_f$ and $2^{-1022}$ for $\delta_f$.
  There is a general trade-off between speed and accuracy,
    though with a lot of heterogeneity between libraries and functions.
}
\label{fig:table}
\end{figure}

Implementations of mathematical functions
  are usually gathered into libraries
  that provide a large collection of functions
  all at a similar point in the speed-accuracy trade-off.
These libraries broadly cover a spectrum
  from faster, less-accurate implementations
  to slower, more-accurate ones.
\name ships with support for many of the most popular libraries,
  covering a range of accuracies and speeds;
  \Cref{fig:table} lists these libraries
  and their error and cost model parameters
  $D$, $\varepsilon_f$, $\delta_f$, and $c_f$.

The golden standard for function implementation are correctly rounded
   implementations.
These yield the true mathematical result, rounded to floating-point;
  this value is unique,
  so all correctly-rounded libraries produce identical answers on all inputs.
Unfortunately, achieving this accuracy is still a topic of active research.
CRLibM provides correctly-rounded implementations
  of many math functions in double precision,
  but it is quite slower, usually by a factor of $5$--$10\times$,
  than a traditional implementation.
Work on reducing this overhead is ongoing.
The recently-published RLibM library
  achieves correct rounding and speed comparable to alternative libraries
  using techniques from program synthesis and verification.
However, its techniques do not scale to 64-bit implementations,
  so RLibM only provides 32-bit implementations
  and furthermore only for a small set of mathematical functions
  not including, for example, \F{sin} and \F{cos}.
Ultimately, neither CRLibM or Rlibm is (currently) in common use, showing that
  most users have already decided on higher speed in exchange for lower
  accuracy.

Standard system math libraries aim for nearly-correct rounding.
These libraries, which include GLibC, OpenLibm, and AMD Libm,
  aim to achieve the highest possible performance
  while allowing the least- or even second-least-significant bit to be incorrect.
Being marginally less accurate than CRLibM and RLibM
  often allows them to be dramatically faster.
Consequently, these libraries are appropriate for general-purpose code,
  where speed is a top-level concern
  but where programmers do not have the expertise
  to optimize the speed-accuracy trade-off further.
Different libraries in this category
  use different implementation strategies
  in order to achieve maximum speed,
  including (often) custom implementations
  for specific hardware architectures or generations.
For example, GLibC has roughly a dozen
  architecture-specific implementations of the main mathematical functions.

In some domains, greater speed and lower accuracy are required;
  for such programs, special-purpose mathematical libraries exist.
For example, the VDT library, developed by CERN,
  allows for up to 3 incorrect bits,
  two more than the standard system library
  (more in single precision).
In exchange for lower accuracy,
  VDT can be up to twice as fast
  as the corresponding GLibC function
  and also has some vectorization advantages.
VDT uses Pad\'e approximations (a variant on polynomial approximations)
  and tunes them to slightly lower accuracy in order to achieve this speedup.

Finally, some applications are best run
  at a speed-accuracy trade-off not represented
  by any of the above libraries.
For such cases, MetaLibm provides a metaprogramming framework
  ideal for generating new function implementations.
MetaLibm provides easy access to polynomial generation with Sollya,
  a convenient cross-compilation mechanism,
  and utility routines useful for writing function implementations.
Of particular interest is that MetaLibm implementations
  can be parameterized;
  for example, MetaLibm's \F{exp} implementation
  can be parameterized by any number of polynomial terms.
Besides the specific implementations that ship with MetaLibm,
  such as the parameterized \F{exp} implementation,
  users can also use MetaLibm to write their own custom implementations.

Some libraries also provide multiple implementations of the same function
  for tuning purposes.
The simplest case of this is providing different implementations
  for single- and double-precision arguments:
  single-precision versions are fundamentally less accurate due to rounding,
  so different parameter choices are advised.
The single-precision versions also usually have a narrow input range.
Usually the single-precision version is not implemented independently;
  instead it uses a truncated and rounded form
  of the polynomial used for the double function.
(This is not optimal, but economizes on the high cost
  of developing a novel function implementation.)
Some libraries also provide function implementations
  valid over only a certain input range,
  which allows those implementations to skip (or simplify) range reduction.
For example, the VDT library internally
  contains implementations of \F{sin} and \F{cos} which are valid
  only on the reduced range $[-\pi/2, \pi/2]$.
These types of implementations are intended to be used with external range
  reduction, and are often not exposed at a library level, but using them
  directly can lead to generous advantages in speed.
The bar to using these implementations
  is proving that the input will be in the reduced range.


\section{Selecting optimal implementations}
\label{sec:optimization}

Between the various libraries listed above,
  and the parameterized implementations from MetaLibm,
  there are dozens of implementations
  of \F{sin}, \F{cos}, \F{tan}, \F{exp}, and \F{log},
  which makes selecting the right one a chore.
\name uses its error and cost models
  to encode the implementation selection problem
  as an integer linear program
  and select the right configurations automatically.

\subsection{Encoding Configuration Selection}

\begin{figure}
\newcommand{\eqnbox}[1]{\fbox{\textsc{#1}}\;}
\begin{mathpar}
\eqnbox{DefVar}
t_{i,j} \in \{ 0, 1 \} \and
\eqnbox{PickOne}
\sum_j t_{i, j} = 1 \and
\eqnbox{SetEps}
\varepsilon_i = \sum_j t_{i,j} \varepsilon_{f_j} \land
\delta_i = \sum_j t_{i,j} \delta_{f_j} \and
\eqnbox{MinErr}
\min \sum_i A_i \varepsilon_i + \sum_i B_i \delta_i \and
\eqnbox{SetCost}
c_i = \sum_j t_{i,j} c_{f_j} \and
\eqnbox{MinCost}
\min \sum_i c_i
\end{mathpar}
\caption{And integer-linear-program formulation of the implementation
  selection problem.}
\label{fig:ilp-form}
\end{figure}

A valid implementation selection
  must minimize both overall error and execution time
  while making a discrete choice of implementation for each use site.
This naturally fits the integer linear programming paradigm;
  a full integer linear program is given in \Cref{fig:ilp-form}.
The key decision variables for the linear program
  are boolean variables that determine
  which implementation is used at each use site:
  $n \times m$ boolean variables $t_{i,j}$,
  where $t_{i,j}$ is true if use site $i$ uses implementation $f_j$
  (\textsc{DefVar}).
Note that for a fixed $i$, the $t_{i,j}$ sum to 1 (\textsc{PickOne}).

The two key constraints on these decision variables
  are to minimize the error and cost models.
First, the $\varepsilon_i$, $\delta_i$, and cost $c_i$
  of the chosen implementation for each use site $i$
  are computed (\textsc{SetEps} and \textsc{SetCost}).
Note that these constraints embed
  the $\varepsilon_f$, $\delta_f$, and $c_f$ constants
  for the available implementations of each function.
Then, the error and cost models are minimized
  in the integer linear program.
The cost model is just the sum of per-use-site costs,
  so is simple to express (\textsc{MinCost}).
The error model, on the other hand,
  is the linear expression described in \Cref{sec:error-and-cost-models}.
In our implementation
  error Taylor series are computed
  with FPTaylor~\cite{fptaylor}
  and each optimization problem is solved
  using the Gelpia~\cite{gelpia}
  sound global nonlinear optimization engine.

To solve these constraints,
  the integer-linear-program solver
  must support two additional features.
The $\varepsilon_{f_j}$ and $\delta_{f_j}$ coefficients
  in this linear program
  typically vary dramatically in magnitude,
  (rounding error can differ dramatically
  between implementations).
Unfortunately, many common ILP solvers
  produce errors or invalid solutions
  when coefficients vary so widely,
  likely due to rounding error inside the solver itself.
An exact rational ILP solver is thus necessary.
Furthermore, the error and cost models
  are usually at odds with one another and
  cannot be minimized simultaneously.
Instead of a single solution to the conflicting goals
  there is a set of points where decreasing error must increase cost,
  and vice versa.
The ILP solver in question must implement
  a Pareto mode which can find this set automatically.
(Repeated ILP queries with a varying error bound also work,
  but an in-solver Pareto mode is typically much more efficient.)
\name uses Z3 as its ILP solver~\cite{z3};
  Z3 uses exact rational arithmetic
  and has an efficient Pareto mode.
A particularly handy aspect of Z3's Pareto modes
  is that the points are generated sequentially, not all at once,
  so users can start exploring \name-suggested configurations
  even while \name is still running.

\subsection{Verification and Timing}

The Pareto-optimal solutions
  to the integer linear program in \Cref{fig:ilp-form}
  are Pareto-optimal for \name's error and cost \textit{models}.
But \textit{modeled} and \textit{real} speed and accuracy,
  differ enough to shift the position of the points.
\name thus recomputes true error and measures actual speed in a post pass.

\name's error model simplifies floating-point error in several ways.
For one, \cref{rdmodel} mildly overestimates floating point error.
Dropping higher-order terms from the error Taylor form
  then underestimates error
  when the $\varepsilon$ and $\delta$ values are large---%
  that is, for particularly inaccurate function implementations.
Then, linearizing the error model in \cref{linearform}
  overapproximates by ignoring correlations between different terms.
So, \name recalculates the accuracy
  of all configurations returned by the ILP solver
  using FPTaylor's most accurate \F{fp-power2} mode.%
\footnote{
Unfortunately, incorporating fixes into the optimization problem
  would require mixed, global optimization
  with both linear and quadratic constraints (MIQP)
  which is too difficult and slow for available solvers.
}
Since the recalculation is working with a fixed implementation selection,
  the higher-order and correlation terms can be included
  and a sound worst-case error bound can be returned.

\name's cost model also oversimplifies.
Needless to say, modern CPUs are complex,
  with features like out-of-order execution and branch prediction
  that \name's cost model ignores.
Runtime also typically depends on the larger application context;
  a table-based function implementation runs much faster
  if the table stays resident in cache.
\name thus measures the speed
  of each configuration returned by the ILP solver.
To make sure the measurement is accurate,
  \name uses the maximum-speed compiler flags,
  averages multiple measurements,
  pre-generates random valid inputs,
  and binds the timing program to a core,
  just like when computing $c_f$ for each implementation.

After accuracies are recalculated and speeds measured,
  it is possible for some configurations returned by the ILP solver
  to no longer be on the Pareto frontier---%
  that is, for one of the returned points
  to in fact be both faster and more accurate than another.
\name filters out such points
  by sorting returned points by actual speed,
  and removing any points whose accuracy is not monotonically increasing.
This filtering is just a convenience for the user,
  and removes, on average, only $26\%$ of the points.
Usually, the points removed by this process are still
  quite close to the Pareto frontier;
  however, removing these points
  means the user needs to explore fewer configurations
  and can find a good configuration more quickly.

\section{Leveraging Restricted Input Ranges}
\label{sec:ranges}

As discussed in \Cref{ssec:libraries},
  there are advantages to reduced range function implementations:
  by skipping range reduction entirely,
  or by using faster but less accurate range reductions,
  a function implementation can be dramatically sped up
  without meaningfully increasing its error.
However, restricted-range implementations can only be safely used
  by proving that its input lies within that restricted range.

\subsection{Modeling Input Ranges}

When a function is called in a floating-point expression,
  the range of inputs it is called on
  depends on not only the mathematical value of the input expression
  but also on how much error that input is computed with.
This dependence means that
  choosing a rougher approximation for one part of a computation
  can change what implementations are available in a later part of the computation.
Determining where a reduced-range implementation can be used
  is therefore like flattening a rug with a bump in it---%
  as soon as one part is fixed another part pushes back up---%
  because replacing one function implementation with another
  will also change the range of its output
  and thus the valid implementations of other function calls.
Yet the advantages of reduced-range implementations
  are too large to pass up.

Formally, consider a sequence of function assignments
  $\tilde{x}_n := \tilde{f_n}(\tilde{x}_{n-a_n}, \tilde{x}_{n-b_n}, \ldots)$,
  except that the function $\tilde{f}_n$ is only valid
  when its first argument is in the range $[d_1, d_2]$
  (functions with range restrictions on other arguments work analogously).
Since $\tilde{f_n}$ is in fact called on $\tilde{x}_{n-a_n}$,
  we must establish that $d_1 < \tilde{x}_{n-a_n} < d_2$.
Following the idea of the error Taylor series,
\[
|\tilde{x}_{n-a_n} - x_{n-a_n}| \leq
\sum_i A^n_i \varepsilon_{f_i} + \sum_i B^n_i \delta_{f_i}
+ o(\varepsilon^2)
\]
  where the $A^n_i$ and $B^n_i$ constants
  are as defined in \Cref{sec:error-and-cost-models}.
For now, ignore the higher-order terms.
Now, suppose $x_{n-a_n}$,
  the true mathematical value of $\tilde{x}_{n - a_n}$,
  is bounded within $[a_1, a_2]$ for all $x \in I$;
  \name computes these bounds using Gelpia.
The requirement $d_1 < \tilde{x}_{n-a_n} < d_2$ can then be rewritten:
\newcommand{\lra}{\operatorname{\Longleftrightarrow}\;}
\begin{align}
 &d_1 < x_{n-a_n} \pm \left( \sum_i A^n_i \varepsilon_{f_i} + \sum_i B^n_i \delta_{f_i} \right)< d_2 \nonumber\\
 \lra& d_1 < a_1 - \left(\sum_i A^n_i \varepsilon_{f_i} + \sum_i B^n_i \delta_{f_i} \right) \land
 a_2 + \left( \sum_i A^n_i \varepsilon_{f_i} + \sum_i B^n_i \delta_{f_i} \right) < d_2 \nonumber \\
 \lra& \left(\sum_i A^n_i \varepsilon_{f_i} + \sum_i B^n_i \delta_{f_i}\right) < a_1-d_1
 \land \left(\sum_i A^n_i \varepsilon_{f_i} + \sum_i B^n_i \delta_{f_i}\right) < d_2 - a_2 \nonumber \\
 \lra& \left(\sum_i A^n_i \varepsilon_{f_i} + \sum_i B^n_i \delta_{f_i}\right) < \min(a_1-d_1, d_2 - a_2)
 \label{inputvalid}
\end{align}
Define the constant $S_{i,j}$
  to be the right hand side value, $\min(a_1 - d_1, d_2 - a_2)$,
  for using implementation $\tilde{f}_j$ at use site $i$.
Crucially, the right hand side value
  depends only on the function input range $[d_1, d_2]$
  and the argument's error-free range $[a_1, a_2]$.
Thus, the final inequality of \Cref{inputvalid}
  is a linear inequality over the same $\varepsilon$ and $\delta$ variables
  as the integer linear program in \Cref{sec:optimization}.
\name adds the following inequality
  to the integer linear program:
\[
\newcommand{\eqnbox}[1]{\fbox{\textsc{#1}}\;}
\eqnbox{InRange} t_{i,j} \implies \sum_k A^i_k \varepsilon_k + \sum_k B^i_k \delta_k < S_{i,j},
\]
The implication can also be written as a pure ILP statement,
  but the ILP solver \name uses, Z3,
  supports implication statements like these directly.
Note that in some cases, the constant $S_{i,j}$ can be negative,
  meaning that $\tilde{f}_j$ cannot be used.

Just like with \name's use of the linear error model,
  a post-pass is necessary
  to take into account the higher-order terms
  that \name ignores during ILP solving.
This post-pass recomputes the input range
  using the most accurate rounding model
  and with higher-order terms bounded using Lagrange's theorem,
  and removes the point from the speed-accuracy Pareto curve
  if adding the higher-order terms makes the chosen implementations invalid.
In practice this occurs for \nBenchDomainViolationRate of points,
  mostly at the far end of the Pareto curve where error is already high.

\subsection{Implementing New Restricted-range Libraries}
\label{ssec:bounding}

To test \name's support for restricted input ranges,
  we used MetaLibm to implement custom restricted-range versions
  of $\sin$, $\cos$, $\tan$, $\exp$, and $\log$.
Our implementations are based on those of \citet{faster-math-functions,even-faster-math-functions}:
  they fit a polynomial of 1--\nMaxTerms terms
  using over a function-specific core interval $I$,
  and then use either no range reduction
  or a custom, simplified range reduction
  without higher-precision arithmetic operations.
Since these range reductions use simplified arithmetic,
  they are accurate over roughly 1--50 multiples of $I$,
  with an accuracy/input range trade-off
  on top of the accuracy/speed trade-off.
Traditionally, restricted-range implementations
  are hidden from library users
  because it is too easy to misuse them.
Because \name is sound, this fear no longer applies.

Implementing these functions
  was relatively straightforward thanks to MetaLibm,
  but deriving accuracy bounds was more challenging.
Inaccuracy---error---comes from three sources:
  algorithmic error, or
  the difference between the polynomial and the function being implemented;
  rounding error, or
  the error of evaluating the polynomial in floating-point arithmetic;
  and reduction error
  from error in the range reduction and reconstruction steps.
To derive a sound bound
  we bound each type of error separately and then sum them,
  once for the $\varepsilon$ parameter
  and once for the $\delta$ parameter.
The algorithmic error is a purely mathematical artifact,
  and Sollya can compute it automatically
  (it is computed in the inner loop of the Remez exchange algorithm).
To bound the rounding error we use FPTaylor
  with the most accurate \F{fp-power2} error model.
While FPTaylor's bounds are very tight they can in rare cases
  be edged out by manual error analysis,
  such as \citet{horner-bound}'s bound
  on the rounding error of a Horner-form polynomial.
To get the best possible bound
  we use the tigher of FPTaylor's and Oliver's bound,
  since both are sound error estimates.
The reduction error, however, is harder to analyze.

Range reduction typically involves a mix
  of integer and floating-point computations,
  so to bound it we use the approach of Lee, Sharma, and Aiken~\citep{verbit}.
In mixed integer-floating point computations
  the integer values are constant
  for some range of floating-point values;
  for example, when rounding a floating-point value to an integer;
  all of the float values in the range $[0.5,1.5]$ round to $1$.
A single mixed-integer-floating-point computation
  is therefore broken down into multiple computations
  only floating-point variables and integer constants,
  where different computations are used for different $x$ values.
Our restricted-range implementations
  compute $k = \lfloor x / \pi \rfloor$;
  this expression for $k$ is monotonic
  so a fixed $k$ values corresponds to an interval of $x$ values.
The error of the remaining floating-point operations
  is computed using FPTaylor,
  and FPTaylor is again used
  to determine how the approximating polynomial modulates that error.
Aggregating across all $k$ values results in the overall reduction error.
Generally, larger $k$ values lead to larger input ranges
  but also greater error.
To account for this, we give a single implementation
  multiple accuracy specifications (as if it were multiple functions)
  with narrower input ranges having lower error.

On the 8-core machine used for this paper,
  verifying the bounds for  the complete collection
  of \nOptunerImpls custom implementations
  of \F{exp}, \F{log}, \F{sin}, \F{cos}, and \F{tan}
  takes approximately \nTotalGenerationTime.
On average, for each implementation,
  generating the approximating polynomial takes \nSollyaAvg,
  verifying its accuracy without range reduction
  takes \nErrorVerrifyAvg,
  and verifying its accuracy with range reduction
  takes an additional \nReductionErrorAvg
  for each $k$ bound.
Importantly,
  though verifying these implementations takes a lot of time,
  it only needs to be done once,
  and the resulting implementations and accuracy bounds
  are distributed with \name.

\section{Evaluation}
\label{sec:evaluation}

We evaluate whether \name
  can automatically select function implementations
  to achieve a trade-off between speed and accuracy.
The data is gathered on a machine
  with an Intel i7-4793K processor and 32GB of DDR3 memory
  running Debian 10.10 (Buster), GCC 8.3.0, GLibC 2.28, Sollya 7.0, and Python 3.7.3.

\subsection{Methodology}
\label{sec:methodology}

\begin{figure*}
 \setlength{\tabcolsep}{2pt}
  \begin{minipage}[t]{0.78\linewidth}
  \begin{tabular}{lll}
    \textbf{Function} $\sin(x)$ &
    \textbf{Function} $\cos(x)$ &
    \textbf{Function} $\tan(x)$ \\

    {\color{orange}$\bullet$}: $D = [-62.83, 64.40]$ &
    {\color{orange}$\bullet$}: $D = [-62.83, 64.40]$ &
    {\color{orange}$\bullet$}: $D = [-39.26, 40.05]$ \\

    {\color{blue}$\bullet$}: $D = [-1.571, 1.571]$ &
    {\color{blue}$\bullet$}: $D = [-1.571, 1.571]$ &
    {\color{blue}$\bullet$}: $D = [-0.1, 0.89]$ \\

    {\color{cyan}$\bullet$}: $D = [-3.142, 3.142]$ &
    {\color{cyan}$\bullet$}: $D = [-3.142, 3.142]$ &
    \\

    \fbox{\includegraphics[width=0.30\linewidth]{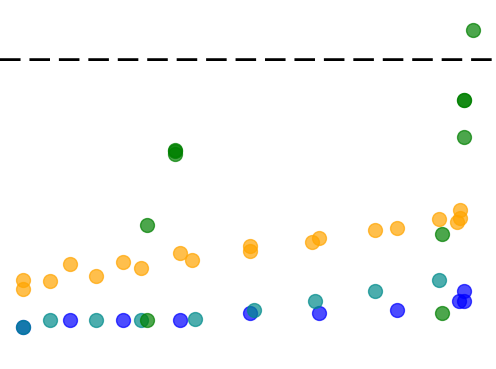}} &
    \fbox{\includegraphics[width=0.30\linewidth]{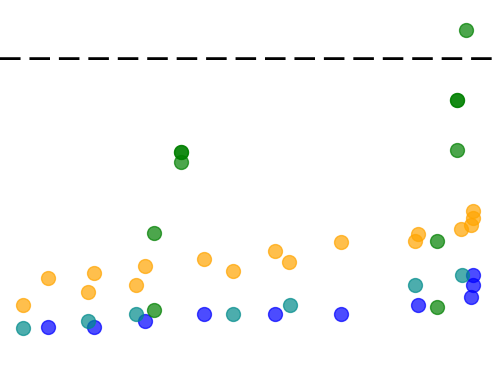}} &
    \fbox{\includegraphics[width=0.30\linewidth]{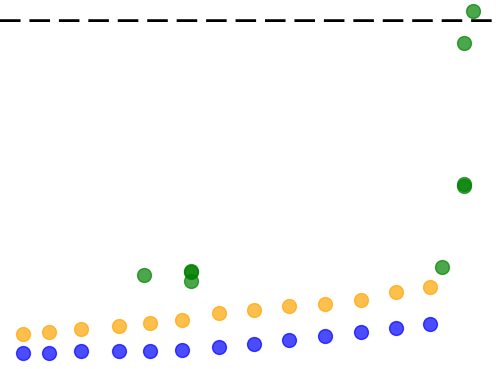}}
  \end{tabular}
  \end{minipage}%
  \hfill%
  \begin{minipage}[t]{0.20\linewidth}
    \scriptsize
    \textbf{Legend} \\[0.025in]
    {\color{green!60!black}$\bullet$}: Standard libraries \\
    {\color{purple}$\bullet$}: MetaLibm \F{exp} \\[0.025in]
    Custom: \\
    {\color{orange}$\bullet$}: mild range reduction \\
    {\color{blue}$\bullet$}: no range reduction \\
    {\color{cyan}$\bullet$}: no range reduction \\
  \end{minipage}

  \vspace{0.5cm}

  \begin{minipage}{0.52\linewidth}
  \begin{tabular}{ll}
    \textbf{Function} $\exp(x)$ &
    \textbf{Function} $\log(x)$ \\

    {\color{orange}$\bullet$}: $D = [-34.75, 35.45]$ &
    {\color{orange}$\bullet$}: $D = [\float{1.5}{-4}, 4096]$ \\

    {\color{blue}$\bullet$}: $D = [-0.1, 0.79]$ &
    {\color{blue}$\bullet$}: $D = [0.65, 1.6]$ \\

    \fbox{\includegraphics[width=0.45\linewidth]{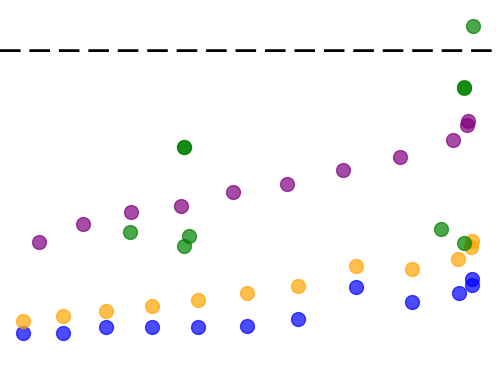}} &
    \fbox{\includegraphics[width=0.45\linewidth]{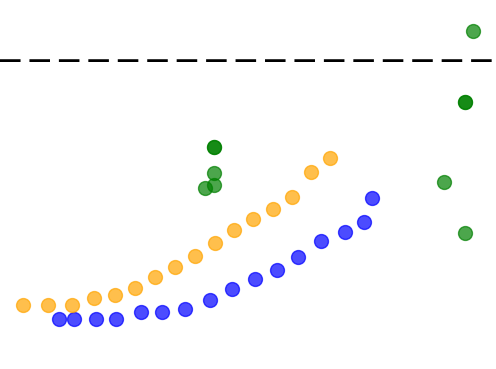}} \\
  \end{tabular}
  \end{minipage}%
  \hfill%
  \begin{minipage}{0.45\linewidth}
  \caption{
    The available implementations for \name.
    Accuracy is shown in log-scale on the horizontal axis,
      while runtime is plotted on the vertical axis.
    The plot's bottom border represents a runtime of 0,
      while the dashed line is a break in the vertical axis
      with CRLibM lying above the break.
    Note that the synthesized implementations
      with simplified range reduction are significantly faster.
  }
  \label{fig:functions}
  \end{minipage}

\end{figure*}

We evaluate \name on \nBenchmarks benchmarks
  from the FPBench suite~\cite{fpbench} as well as
  the \F{haskell} benchmark suite from Herbie 1.5~\cite{herbie},
  originally extracted from Haskell packages via a compiler plugin.
Specifically, we select all benchmarks
  that use the \F{exp}, \F{log}, \F{sin}, \F{cos}, or \F{tan} functions
  and do not contain loops or tensors (which \name does not support).
Within these benchmarks \F{exp}, \F{log}, \F{sin}, \F{cos}, and \F{tan} are
  used a collective \nBenchmarkUseSites times;
  the only other library function called is a single use of \F{atan},
  for which we use the GLibC implementation.
Thirteen benchmarks have a single input, ten have two inputs, eleven have three
  inputs, and the remainder have four or more inputs.
Some of the benchmarks come equipped with input ranges defined for them,
  but, for those that did not,
  we choose input ranges that avoid domain errors such as division by zero.

\name tunes these benchmarks using a total
  of \nOptunerImpls implementations
  of library functions \F{exp}, \F{log}, \F{sin}, \F{cos}, and \F{tan}
  drawn from
  the standard mathematical libraries
  and parameterized MetaLibm implementations
  described in \Cref{ssec:libraries}.
\Cref{fig:functions} plots the accuracies and costs
  for standard mathematical libraries (in green);
  MetaLibm's parameterized \F{exp} implementation (in purple);
  and \name's custom, range-restricted implementations
  (in blue, cyan, and orange).
The various implementations span
  from half an ulp of error to a relative error of $10\%$,
  and the fastest implementation
  is usually about $5\times$ faster than GLibC
  while the slowest is usually about $5\times$ slower.
In between these extremes,
  there is a smooth trade-off between speed and accuracy.

\subsection{Results}

\begin{figure}
  \begin{subfigure}{.5\linewidth}
    \includegraphics[width=\linewidth]{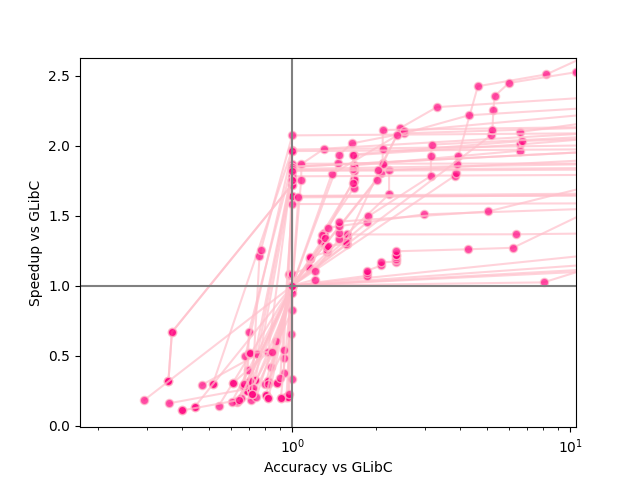}
    \caption{High-accuracy configurations}
    \label{fig:aggregatezoom}
  \end{subfigure}%
  \begin{subfigure}{.5\linewidth}
    \includegraphics[width=\linewidth]{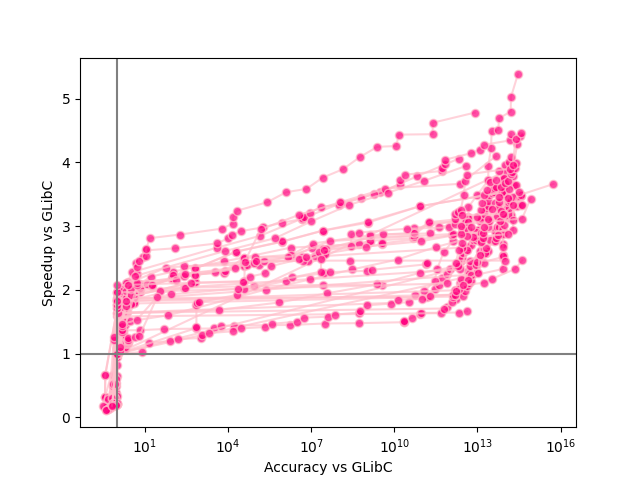}
    \caption{All configurations of any accuracy}
    \label{fig:aggregate}
  \end{subfigure}
    \caption{
      \name's optimized implementations of \nBenchmarks benchmarks.
      Each point's accuracy and runtime is normalized
      so that using the GLibC implementations for each function
      gives a speedup and relative error of $1\times$.}
    \label{fig:bothaggregates}
\end{figure}

\Cref{fig:bothaggregates} contains all \nBenchmarks
  speed-accuracy Pareto curves,
  with \nBenchImpls configurations in total.
In the plot, each line represents a single benchmark,
  and each heavy dot along that line
  is a configuration produced by \name.
The benchmarks are
  normalized so that the ``standard'' implementation
  that uses GlibC for each use site is at $(1, 1)$.

On the left, in \Cref{fig:aggregatezoom},
  are all benchmark implementations
  with error no more than $10\times$ larger
  than the standard implementation.
Note that even with just one decimal digit more error,
  speedups of $50\%$, $100\%$
  and sometimes even $150\%$ are possible.
Note also the cluster of points
  below and to the left of $(1, 1)$.
These points represent configurations that are
  more accurate but also slower than GLibC,
  generally using the correctly-rounded CRLibM functions.
A few points in the plot are above and to the left of $(1, 1)$:
  these configurations are both faster and more accurate than GLibC,
  generally by mixing CRLibM and custom implementations.
In this case, \name is truly offering speed for free.
All told, \Cref{fig:aggregatezoom} shows
  than \name can produce impressive speedups
  with minimal expertise or knowledge of numerical analysis.

The right-hand plot, in \Cref{fig:aggregate},
  instead focuses on applications tolerant of significant error,
  such as POV-Ray.
Here, implementations with dramatically higher error are plotted,
  and correspondingly larger speedups are achieved.
The points on this figure generally use
  \name's custom implementations,
  which are the fastest ones available.
The speedups here are as large as \nBenchmarkLargeSpeedup.
However, the available speedup is limited
  in benchmarks that use mathematical operations
  that \name does not tune
  and in benchmarks with few use sites.
For an average benchmark, therefore,
  the maximum speedup reaches \nBenchmarkAvgMaxSpeedup.
Of course, by the logic of the Pareto curve,
  this maximum speedup comes with minimal accuracy,
  and not all applications are as error-tolerant as POV-Ray.
Nevertheless, the figure shows that most benchmarks' Pareto curves
  feature a steady upward slope,
  meaning that
  decreasing accuracy consistently buys increasing speed.
Many applications could fruitfully use \name
  to explore the possibilities that this trade-off offers.

On most benchmarks \name runs in a few minutes.
\name's run time depends on several factors:
  the number of use sites;
  the complexity of the expression;
  the number of input arguments;
  and in a few cases the input range used.
On our test machine,
  \nBenchmarksUnderThreeMinutes of the \nBenchmarks benchmarks complete in under three minutes
  and \nBenchmarksUnderTenMinutes complete in under ten.
The remaining four benchmarks contain the most use sites and thus have
  the most output configurations to verify and time.
\name generates output configurations incrementally,
  so for these slowest benchmarks,
  users would have \name's first configurations available much sooner,
  usually a few minutes in.

\subsection{Detailed Analysis}

A close inspection of the \name's selected configurations
  demonstrates that these speedups often come from noticing
  use sites that have little impact on accuracy.
Consider the benchmark \F{problem\_3\_3\_2},
  originally from a mathematical textbook~\cite{nmse}:
\[
\begin{array}{l}
\K{require}\: 0 \le \mathit{x} \le 0.75 \land 0 \le \mathit{eps} \le 0.0078125\\
\K{return}\: \tan(\mathit{x} + \mathit{eps}) - \tan(\mathit{x})
\end{array}
\]
\name mixes GLibC's \F{tan} for the first call and VDT's \F{tan} for the second
  to give a \nDetailedOneSpeedup speedup over pure GLibC while only increasing
  relative error from \nDetailedOneErrorGlibc to \nDetailedOneErrorChosen.
The first call to \F{tan} has a larger argument than the second
  call to \F{tan}, and due to \F{tan}'s steep increase in the region just to the
  right of $0$ this gives the first call a higher impact on overall accuracy.
\name notices this and uses a more accurate implementation for
  the first call than the second call.

\name's optimizations can be even more subtle;
  consider this complex sine benchmark:
\[
\begin{array}{l}
\K{require}\:0 \le \mathit{re} \le 10 \land 0 \le \mathit{im} \le 10\\
\K{return}\: (0.5 \cdot \sin(\mathit{re})) \cdot (\exp(-\mathit{im}) - \exp(\mathit{im}))
\end{array}
\]
The second call to \F{exp} requires more accuracy than the first
  because $\mathit{im}$ is positive,
  so the first \F{exp} returns smaller values than the second
  and thus has less impact on the expression's error.
Meanwhile, the output of the \F{sin} call
  is multiplied $\exp(-\mathit{im}) + \exp(\mathit{im})$
  and its error is thus magnified.
In view of these effects,
  \name selects a high accuracy \F{sin} (such as CRLibM's)
  and two different \F{exp} implementations
  (say, order-9 and order-12 custom implementations)
  along the speed-accuracy Pareto curve.
This unintuitive mix of implementations leads to a \nDetailedTwoSpeedup speedup
  over an all GlibC configuration while only increasing error from
  \nDetailedTwoErrorGlibc to \nDetailedTwoErrorChosen.

These same patterns occur far along the speed-accuracy Pareto curve.
Consider the \F{logexp} benchmark:
\[
\begin{array}{l}
\K{require}\:0 \le x \le 8 \\
\K{return}\: \log(1 + \exp (x))
\end{array}
\]
At lower accuracies \name will select
  VDT's float variation of \F{log}
  and a custom order-7 \F{exp} implementation.
This yields a speedup of \nDetailedThreeSpeedup,
  paid for an accuracy drop from \nDetailedThreeErrorGlibc to \nDetailedThreeErrorChosen.
As in all these examples, the most important thing to note
  is the extensive expertise and time commitment necessary
  to do a similar optimizations by hand.


\section{Related Work}
\label{sec:relwork}






Implementing mathematical functions in floating point has a long history.
\citet{kahan, kahan-log}, \citet{higham},
  \citet{muller}, and \citet{cody-waite}
  all made monumental contributions to the field.
These authors all leveraged earlier work on approximation theory
  developed by mathematicians like Pafnutiy Chebyshev
  and Charles-Jean de la Vall\'ee Poussin.
Robin Green's recent talks on the topic~[\citeyear{faster-math-functions,even-faster-math-functions}]
  are a good introduction to and survey of the field.
Standard library implementations tend to be accurate
  to within 1 or a few ulps~\cite{glibc-accuracy},
  but some implementations have been produced
  with a ``gold-standard'' accuracy of  a half ulp,
  including MPFR~\cite{mpfr} and CRLibm~\cite{crlibm}.
Most implementations of library functions like \F{exp} or \F{sin}
  are manually verified, and bugs are sometimes discovered~\cite{intel-sin,faster-math-functions}.
However, a few of the Intel Math Kernel Library implementations
  of functions like \F{log}, \F{sin}, and \F{tan} have been verified
  with semi-automated methods~\cite{verified-math-h},
  and some verified synthesis techniques are available~\cite{jay-p-lim}.
\name could these libraries or any other libraries.

Using lower-accuracy library function implementations
  is an established, if infrequent, program optimization technique.
The CERN math library
  used in \name, VDT \cite{cern},
   was created to allow developers to manually tune this tradeoff.
VDT is also carefully optimized for SIMD support,
  which \name does not currently attempt.
There are other similar libraries that \name does not yet incorporate
  such as
  Intel MKL~\cite{mkl}, CEPHES~\cite{cephes},
  and VC's math functions~\cite{vc}.

\citet{daisy-libm} have attempted to automate
  the implementation selection problem
  in the Daisy numerical compiler~\cite{daisy}.
In a certain sense,
  these authors approach a similar problem to \name, but in reverse.
In their approach,
  Daisy starts by analyzing the user's expression
  to derive an accuracy bound for each call to a library function.
Daisy then leverages MetaLibm's
  parametric implementations~\cite{metalibm}
  to generate a custom implementation for this accuracy bound.
We experimented with using MetaLibm in a similar way.
However, we ultimately found that enumerating the space
  of possible library function implementations was a superior approach,
  not only because it searches over
  a broader range of implementation choices
  than MetaLibm's parameterized implementations
  but also because it allows for much more accurate error estimation
  and thus greater speedups.
Daisy also returns a single implementation,
  while \name returns the full speed-accuracy Pareto curve.

While not specific to floating point, the Green framework~\cite{green-frame}
  does allow replacing math functions with faster variations in the spirit of
  approximate computing.
Instead of relying on error analysis, the code under test must be accompanied
  by a quality of service metric.
Green provides statistical metrics but does not guarantee worst case behavior.
Another downside of this approach is that it scales linearly
  in the number of possible implementation configurations,
  which itself grows combinatorially in the number of available implementations and call sites.

Several tools attempt to
  speeding up floating point computation
  by computing intermediate values to lower precision.
Precimonious~\cite{precimonious} approaches the problem by sampling points and
  dynamically testing the speed and error while lowering precision of
  intermediates in a random search.
Blame Analysis~\cite{prec-blame}
  instead dynamically determines which intermediates have low impact on error,
  and then select those intermediates as candidates for lowering.
HiFPTuner~\cite{hifptuner} improves further upon this method
  by performing a static analysis of the expression
  to group intermediates and hierarchically search the space of precisions.
CRAFT~\cite{lam} is similar to this line of work,
  but performs the dynamic analysis at a binary level.
None of these tools can guarantee a sound error bound.
FPTuner~\cite{fptuner}, on the other hand, performs precision tuning
   while also guaranteeing an overall error bound.
Its use of error Taylor forms and integer linear programming
  was an inspiration for \name.
Optimizing value precision is orthogonal to \name's purpose,
  and could potentially integrated into \name.

Other tools improve the accuracy of a floating point expression without an explicit
  concern for speed.
The Herbie~\cite{herbie} and Salsa~\cite{salsa} tools
  attempt to increase the accuracy of a floating-point expression
  by rewriting it using algebraic and analytic identities.
While these tools do not consider speed,
  they do sometimes discover rewrites
  that both increase accuracy and improve runtime~\cite{herbie}.
More recently, the Herbie authors have added support
  for combining rewriting with precision tuning
  to explore the speed-accuracy trade-off of lower-precision arithmetic~\cite{pherbie}.
Finally, the \textsc{Stoke} tool uses stochastic search over assembly instructions
  to improve runtime without much reducing accuracy~\cite{stoke-fp}.
While this sometimes discovers valuable algebraic rearrangements,
  this approach cannot implement a complex library function like \F{exp} or \F{sin}.
\textsc{Stoke} also cannot bound the worst-case error
  of its tuned floating-point expressions;
  it can only bound the difference between the original and optimized version.
In any case, none of these tools currently considers
  changing the implementation of library functions as \name does.
Integrating \name with these tools
  would likely discover new speedups and further refine
  the speed-accuracy Pareto curves discovered.

\section{Conclusion}
\label{sec:conclusion}

\name makes floating-point computations faster
  by choosing the right implementation
  of library functions like \F{exp} or \F{sin}.
\name uses a linear error model
  and integer linear programming
  to select the best implementations to use
  (from among standard mathematical libraries
  and generated implementations)
  for each use of a library function in a floating-point expression.
It then verifies the optimal configurations
  and presents the user with a speed-accuracy Pareto curve
  summarizing the available optimization options.
Across \nBenchmarks benchmarks,
  \name demonstrates speedups of up to \nBenchmarkLargeSpeedup,
  with speedups of \nBenchmarkSafeSpeedup available
  at negligible accuracy increases.
\name thus demonstrates the possibilities
  of a hitherto-underexplored avenue
  for floating-point program optimization.

\appendix

\section{Appendix: Case Study Methodology}
\label{sec:cs-methods}

To evaluate POV-Ray's speed and accuracy
  for different function implementation configurations,
  we leveraged the SPEC 2017 benchmark suite,
  which includes POV-Ray.
SPEC 2017 includes a standard compilation harness
  and a standard quality measure,
  the structural similarity index measure~\cite{ssim},
  which assigns each $8\times8$ pixel block a score from $-1$ to $+1$,
  with $+1$ indicating an exact match.
The minimum block score is then used to evaluate image quality,
  with scores over \nSpecThreshold considered acceptable.
The benchmark scene used by SPEC does not use the photons feature,
  but POV-Ray ships many standard scenes to demonstrate its capabilities,
  and we use one of those, \F{grenadine},
  to avoid creating our own contrived example.

The version of POV-Ray included in SPEC 2017
  includes and uses the custom \F{sin} and \F{cos} implementations
  described in \Cref{sec:case_study}.
These implementations are quite simplistic;
  a condensed version of \F{sin} is shown in \Cref{fig:table-based}.
The input $x$ is converted to an integer from $0$ to $254$
  by simple linear transformation:
  $i = (x / \pi + 1) * 127$, rounded down.
Then a hard-coded table is used to look up $\sin((i - 127)*\pi / 127) \pi$.
This implementation is quite inaccurate---%
  $i$ only has 8 bits of information---%
  but is also quite fast, likely because during a tight loop
  the hard-coded table stays loaded in cache.
\name's fastest implementations, meanwhile, are all polynomial-based.
Nonetheless, some of the configurations it finds
  are both faster and more accurate than the POV-Ray implementations,
  suggesting that \name could have helped the POV-Ray developers
  speed up caustics.

One challenge in this evaluation is the choice of baseline.
SPEC uses POV-Ray's custom table-based implementations
  to generate its reference image,
  but as mentioned above those implementations are inaccurate.
We therefore modified POV-Ray
  to use the \F{sin} and \F{cos} implementations from GLibC,
  which POV-Ray had historically used
  (prior to the development of the custom table-based implementations),
  and used that implementation
  to generate our reference images as well as our performance baseline.
We also verified
  that a rendering using CRLibM's~\cite{crlibm} \F{sin} and \F{cos}
  produces the exact same image;
  since CRLibM is maximally accurate,
  this further justifies the use of GLibC as a baseline.
Code comments in the POV-Ray source code
  confirm that before POV-Ray added
  its custom \F{sin} and \F{cos} implementations,
  it used GLibC (or other system math library) implementations,
  making this baseline historically plausible.

To use \name with POV-Ray,
   we first extracted the expression seen in \Cref{povprog}
   and recorded it in FPCore, \name's input language.
The bounds on the input variables were easily established
  since the angle values were already known to be bounded
  and the $n$ values come from a unit normal.
We then ran \name to generate the configurations
  along the speed-accuracy Pareto curve.
For each resulting configuration,
  we produced a patched version of POV-Ray
  by injecting the source code
  for all of \name's supported implementations into POV-Ray
  and using macros and compiler flags to select which one is used.
The SPEC harness was then used
  to determine both the speedup and the quality of the resulting rendering.
This allows us to measure \name's outputs
  not just in terms of the isolated extracted expression
  but also in end-to-end real terms
  for a large and performance-sensitive software project.

Besides the change in function implementations,
  the POV-Ray developers made a second change
  to the photon incidence computation,
  again trading away accuracy to gain speed,
  but in this case by modifying the storage format
  instead of by changing the implementation of functions.
Specifically, instead of storing incidence angles as double-precision values,
  as prior POV-Ray releases and commented-out code did,
  they changed some internal data structures
  to store the index $i$ directly.
Since indices can be stored in a single byte,
  this reduces the memory used for the photon table%
\footnote{It also seems to reduce padding and alignment issues,
  since double-precision values have to be 8-byte aligned on our system.}
  which speeds POV-Ray up by a further \nPovRayCharSpeedup,
  resulting in the speed and accuracy shown by the orange circle
  in \Cref{fig:povray-end2end}.
This is faster than the best speed-up achievable
  by tuning function implementations alone.
\name only supports double-precision computation
  and cannot make storage optimizations,
  so it cannot directly propose a similar optimization.
That said, we did make some ad-hoc modifications to \name
  to generate error models for 8-bit inputs
  and to produce a speed-accuracy Pareto curve with them.
\name is again able to find configurations
  both faster and more accurate than that used in POV-Ray.
This is possible because POV-Ray's implementation
  is still inaccurate, even for 8-bit inputs,
  because when building the table it evaluates \F{sin}
  at the location corresponding to index $i$,
  instead of the location corresponding to $i + \frac12$.
These results are tentative:
  our ``8-bit'' function implementations are just
  wrapped versions of our double-precision implementations,
  and \name's support for 8-bit values is limited.
That said, these results suggest that
  precision tuning along the lines of FPTuner~\cite{fptuner} or POP~\cite{pop}
  could be combined with \name's tuning of function implementations
  to suggest even faster and more accurate configurations.

\bibliography{bibliography}

\end{document}